\documentclass[12pt]{article}
\usepackage[left=2.5cm,right=2.5cm,top=2.5cm,bottom=2.5cm]{geometry}
\usepackage{amsfonts,amssymb}
\usepackage[utf8]{inputenc}
\usepackage{xcolor}
\usepackage{amsmath}
\usepackage{esvect}
\usepackage{framed}
\colorlet{shadecolor}{lightgray}
\usepackage{hyperref}
\usepackage{cite}
\usepackage{parskip}
\usepackage{graphicx}
\usepackage{float}

\begin{document} 

\baselineskip 18pt

\begin{center}

{\Large \bf Dynamical analogue spacetimes in non-relativistic flows}

\end{center}

\vskip .6cm
\medskip

\vspace*{2.0ex}

\baselineskip=18pt

\centerline{\large \rm Karan Fernandes, Susovan Maity and Tapas K. Das}

\vspace*{4.0ex}

\centerline{\large \it ~Harish-Chandra Research Institute,}
\centerline{\large \it  Chhatnag Road, Jhusi, Prayagraj 211019, India.}

\vspace*{1.0ex}
\centerline{\small E-mail: karanfernandes@hri.res.in, susovanmaity@hri.res.in, tapas@hri.res.in}

\vspace*{5.0ex}

\centerline{\bf Abstract} \bigskip

Analogue gravity models describe linear fluctuations of fluids as a massless scalar field propagating on stationary acoustic spacetimes constructed from the background flow. 
In this paper, we establish that this paradigm generalizes to arbitrary order nonlinear perturbations propagating on dynamical
analogue spacetimes. Our results hold for all inviscid, spherically symmetric and barotropic non-relativistic flows in the presence of an external conservative force. We demonstrate that such fluids always admit a dynamical description governed by a coupled pair of wave and continuity equations. We provide an iterative approach to solve these equations about any known stationary solution to all orders in perturbation. In the process, we reveal that there exists a dynamical acoustic spacetime on which fluctuations of the mass accretion rate propagate. The dynamical acoustic spacetime is shown to have a well defined causal structure and curvature. In addition, we find a classical fluctuation relation for the acoustic horizon of the spacetime that admit scenarios wherein the horizon can grow as well as recede, with the latter being a result with no known analogue in black holes. 
As an example, we numerically investigate the Bondi flow accreting solution subject to exponentially damped time dependent perturbations. 
We find that second and higher order classical perturbations possess an acoustic horizon that oscillates and changes to a new stable size at late times. In particular, the case of a receding acoustic horizon is realized through `low frequency' perturbations. We discuss our results in the context of more general analogue models and its potential implications on astrophysical accretion flows.

\vfill \eject

\baselineskip 18pt

\tableofcontents

\section{Introduction}

Linear fluctuations of transonic fluids can be described in terms of a massless scalar field propagating on a relativistic spacetime containing acoustic horizon(s) from within which phonons cannot escape~\cite{Moncrief: 1980,Unruh:1980cg,Visser:1993ub, Visser:1997ux,Bilic:1999sq,Novello:2002qg,us:07,Barcelo:2005fc}. This result forms the basis of the analogue gravity programme, to investigate properties of fields on black hole spacetimes through analogues of fluctuations on acoustic spacetimes~\cite{Novello:2002qg,us:07,Barcelo:2005fc,Faccio:2013kpa}. 
While originally motivated in the context of classical fluids, 
analogue gravity models are now known to exist in a much broader context -- from superfluids such as Bose-Einstein condensates~\cite{Garay:1999sk,Garay:2000jj,Barcelo:2000tg,Barcelo:2003wu} and superfluid helium~\cite{Jacobson:1998he,Volovik:1999fc,Volovik:2000ua,Volovik:2002ci} to quantum optics and electromagnetic waves in dielectric media~\cite{Corley:1998rk,Leonhardt:2000fd, Leonhardt:2000zp,Schutzhold:2004tv,Philbin:2007ji}, to name a few.

Analogue models were primarily motivated to experimentally test aspects of black hole evaporation~\cite{Unruh:1980cg,Visser:1993ub,Visser:1997ux}, particularly Hawking radiation and the trans-Planckian problem~\cite{Jacobson:1991gr,Corley:1996ar,Unruh:1994je,Unruh:2004zk}. Over the past decade, there have been several important experimental results including the stimulated Hawking effect~\cite{Weinfurtner:2010nu,Steinhauer:2014dra,Drori:2018ivu}, spontaneous emission in Bose-Einstein condensates and fibre-optic set-ups~\cite{Lahav:2009wx,deNova:2018rld}, entanglement in analogue Hawking radiation~\cite{Steinhauer:2015saa}, and analogues of quasi-normal modes~\cite{Torres:2018dso} and superradiance~\cite{Torres:2016iee} in the draining bathtub model. These observations follow from high frequency linear fluctuations about static or stationary analogue spacetimes. More recent experiments have also probed aspects of dynamical analogue spacetimes. Among these results are those for the spontaneous Hawking radiation through the evaporation of analogue black holes in Bose-Einstein condensates~\cite{Kolobov:2021} and the time dependent behaviour of analogue spacetimes in the draining bath tub model following higher order fluctuations~\cite{Goodhew:2019tax}. 
These observations demonstrate that the time dependence involved in higher order fluctuations of analogue spacetimes are relevant and can generically grow to have observable effects. 
In addition, as nonlinear fluctuations propagate on dynamical analogue spacetimes, such experiments could in general provide insights on dynamically evolving black hole spacetimes.

There currently are a number of back-reaction results~\cite{Goodhew:2019tax,Balbinot:2004da,Schutzhold:2005ex,Balbinot:2006ua,Balbinot:2006cy,Maia:2007vy,Liberati:2020mdr} for second order fluctuations propagating on dynamical analogue spacetimes. These results are quasi-stationary, arising from certain mean-field approximations needed to characterize the compressibility of a given unperturbed incompressible fluid. 
A more complete description of dynamical analogue spacetimes and their nonlinear properties will require perturbations to all orders, which remains an open problem in the field. 
The primary difficulty in this derivation concerns the absence of dynamical equations for the analogue metric. 
In considering transonic flows, a steady state solution is typically needed to characterize the fluid. The analogue spacetime arises as an emergent construct at each order in perturbation from recasting the perturbed equations for the fluctuation as a wave equation involving an acoustic spacetime metric.

An important example of compressible transonic fluids are astrophysical accretion flows, which naturally involve a changing mass density over time. Perturbations of fluids that accrete into black holes are particularly interesting as they uniquely propagate on analogue spacetimes that include both event horizons and acoustic horizons in their description. 
Linear perturbations of accreting astrophysical systems provide classical analogue gravity models that have been well studied in the past. Moncrief~\cite{Moncrief: 1980} established that spherically symmetric accretion flows on a Schwarzschild spacetime are stable under linear perturbations of the velocity potential, which propagate on a static analogue spacetime. Analogue spacetimes in astrophysical accretion flows are now known to arise more generally in axisymmetric flows~\cite{Das:2003py,Abraham:2005ah,Das:2006an,Bilic:2012yh,Tarafdar:2013oqa}, isothermal flows~\cite{Shaikh:2016qko,Shaikh:2017mfs}, flows under post-Newtonian potentials~\cite{Dasgupta:2005hh} and on Kerr spacetimes~\cite{Pu:2012rv,Saha:2015lka,Tarafdar:2017jts,Shaikh:2018kiq}. Apart from velocity perturbations, analogue spacetimes in accreting fluids also follow from linear perturbations of the mass accretion rate~\cite{Naskar:2007su,Ananda:2014gga,Ananda:2014qba} and the Bernoulli constant~\cite{Datta:2016bgm}. Such analogue spacetimes might provide insights on observables in the supersonic regions of accretion flows, as flow variables and their terminal values can be determined from pertubations close to the acoustic horizons.

The fluctuating field of transonic fluids is usually considered to be that of the velocity potential. In this case, the Euler equation of a non-relativistic fluid can be cast as the Schr\"odinger equation, with applications to the Hawking effect on analogue spacetimes. However, the velocity potential is a relevant variable only when the fluid is irrotational. 
The mass accretion rate on the other hand is a quantity that can be defined for all fluids, regardless of the irrotational condition. It characterizes the compressibility of the flow and takes on constant values for a stationary flow. As such, the fluctuation of the mass accretion rate could be of particular relevance in dynamical analogue spacetimes. It is also a variable of fundamental importance in astrophysical accretion flows due to its direct relationship with observables.  
A formalism that accounts for nonlinear perturbations of the mass accretion rate could hence be used to probe dynamical astrophysical accretion flows in their supersonic regimes.

In this paper, we establish that dynamical analogue spacetimes arise from arbitrary order nonlinear perturbations for a certain class of fluids. We specifically consider inviscid, irrotational and barotropic non-relativistic flows in the presence of a conservative force.   
Our analysis is based on the choice of the mass accretion rate and the density of the fluid as independent variables. We demonstrate that this choice allows for the general dynamics of the fluid to be determined from two coupled equations -- a continuity equation relating the mass accretion rate and density, and a wave equation for the mass accretion rate. These equations can be solved perturbatively about any stationary solution to all orders using an iterative approach. The resulting expressions at each order in perturbation identify a mass accretion rate propagating on an effective dynamical analogue spacetime. The dynamical acoustic metric is shown to have well defined curvature and causal properties to all orders in perturbation. We additionally consider the relative variation of the acoustic horizon radius between two orders of perturbation. Unlike the event horizons of black holes which always grow under classical perturbations, we identify cases where the acoustic horizon could recede depending on the relative variation of the density and mass accretion rate perturbations. 

To manifest our formalism, we numerically investigate exponentially damped time dependent perturbations about the transonic Bondi accretion flow solution~\cite{Bondi:1952}.
There exist two relevant length scales in this problem, namely the outer radius of the accreting fluid and the boundary of the accreting body, taken as the inner radius of the flow. The existence of two length scales allow for time dependent perturbations in broadly two frequency regimes which we label as `high frequency' and `low frequency'.  Perturbations with wavelengths less than the size of the accretor are considered as `high frequency', while perturbations with wavelengths greater than the size of the accreting fluid are termed `low frequency'. 
To linear order in perturbation, we recover the known static analogue spacetime which is the zeroth order effective analogue spacetime in our framework. 
At higher orders in perturbation, the acoustic spacetime involves an oscillating horizon that settles to a new radius at late times. 

In the high frequency case, the second order mass accretion rate perturbation propagates on a first order effective acoustic spacetime with a sonic horizon that oscillates and grows to a new radius. This behaviour is analogous to the ringdown of black holes under classical perturbations. However in the low frequency case, the first order effective analogue spacetime involves an oscillating horizon that recedes to a new radius at late times. This result has no known analogue in black holes and could be used as a novel experimental test for analogue spacetimes that we introduce in this work. Low frequency perturbations beyond second order are qualitatively similar to the high frequency results, that provide a relative growth of the acoustic horizon. We demonstrate however that these higher order corrections are suppresed by the perturbation parameter regardless of the strength of the perturbation. Hence the first order effective acoustic spacetime in all frequency regimes provide a good approximation to the nonlinear dynamics of the acoustic horizon.   

Our paper is organized as follows. In Sec.~\ref{nrf} we describe the spherically symmetric fluid and our perturbation scheme leading to the result of an effective acoustic spacetime on which arbitrary order nonlinear perturbations  of the mass accretion rate propagate. In Sec.~\ref{curve}, we discuss the geometry of the analogue spacetime, demonstrating that it possesses a well defined curvature and causal structure to all orders in perturbation. We further provide a fluctuation relation for the acoustic horizon and discuss the scenario under which it can shrink under perturbations. In Sec.~\ref{num}, we numerically determine results up to the second order effective acoustic metric resulting from an exponentially damped in time perturbation of the Bondi solution. We discuss in detail the assumptions involved in the high and low frequency cases along with subsequent results for the acoustic horizon. We conclude with a discussion on some future directions.

\section{Nonlinear perturbations of spherically symmetric flows} \label{nrf}

The system we will consider is that of an inviscid, spherically symmetric and barotropic non-relativistic fluid in the presence of a conservative force. Given the spherical symmetry of the problem, the fields can depend only on the radial distance `$r$' and time `$t$'. We will denote the density and pressure of the fluid by $\rho$ and $P$ respectively. The potential of the external force depends only on $r$ and will be denoted by $\Phi$. The velocity vector of the fluid has only one non-vanishing radial component which we denote by $v$. With these assumptions, the dynamics of the fluid are governed by the continuity equation
\begin{equation}
\dot{\rho} + \frac{1}{r^2}\left(r^2\rho v\right)' = 0 \,, \label{cont}
\end{equation}
and the Euler equation
\begin{equation}
\dot{v} + v\,v' + \frac{1}{\rho} P' + \Phi' = 0 \, \label{eul1}\,,
\end{equation}
where overdots denote time derivatives and primes spatial derivatives.  To complete the description of the fluid we need to specify an equation of state. We assume the fluid to be barotropic which allows us to define a scalar function $H(\rho)$ such that $H' = \frac{1}{\rho} P'$. Barotropic fluids accomodate a variety of special cases including the isentropic and isothermal conditions. If we define the fluid sound speed $c_s$ by $c_s^2 = \frac{\partial P}{\partial\rho}$, then we can determine the relation
\begin{equation}
\frac{\partial H}{\partial \rho} = \frac{c_s^2}{\rho}\,.
\label{ss}
\end{equation}

With the barotropic condition, we can express Eq.~(\ref{eul1}) as
\begin{equation}
\dot{v} + \left(\frac{1}{2}v^2 + H + \Phi\right)' = 0
\label{eul}
\end{equation}

Linear perturbations of Eq.~(\ref{cont}) and Eq.~(\ref{eul}) are considered by first expanding the fields around some stationary solution. The perturbed field of interest for irrotational flows is typically taken to be the velocity potential $\Psi$, where $v = \Psi'$. The time derivative of the perturbed Euler equation can then be cast into a wave equation for $\Psi$ propagating on a stationary analogue spacetime constructed from the background flow. The generalization of this procedure to higher orders would involve expanding the fields to the respective order, followed by re-expressing the time derivative of the perturbed Euler equation as a wave equation. This procedure gets increasingly involved with each order, with no guarantee of a resulting equation for a massless scalar field on a curved analogue spacetime.

In the following, we will proceed with a different scheme. Our approach is based on the mass accretion rate, defined in the spherically symmetric case by
\begin{equation}
f = \rho v r^2 \,.
\label{mac}
\end{equation}
We will now express Eq.~(\ref{cont}) and Eq.~(\ref{eul}) entirely in terms of $f$ and $\rho$.
Using the definition in Eq.~(\ref{mac}), we find that Eq.~(\ref{cont}) can be readily expressed as
\begin{equation}
\dot{\rho} + \frac{1}{r^2}f' = 0 \,.
\label{cont2}
\end{equation}
This is the continuity equation in our perturbation scheme. We will now demonstrate that we can derive a wave equation directly from the Euler equation prior to any perturbation. 
Using the expressions in Eq.~(\ref{cont2}) and Eq.~(\ref{ss}), we find that the time derivative of Eq.~(\ref{eul}) takes the form
\begin{equation}
\ddot{v} + \left( v \dot{v} - \frac{c_s^2}{r^2 \rho} f' \right) ' = 0
\label{teul}
\end{equation}

Taking the time derivative of Eq.~(\ref{mac}) and using Eq.~(\ref{cont2}), we find the following expression for $\dot{v}$ entirely in terms of derivatives of $f$ and $\rho$
\begin{equation}
\dot{v} = \frac{1}{\rho r^2}\left( \dot{f} + \frac{f \partial_r f}{\rho r^2} \right)
\label{vdot} 
\end{equation}

We can now substitute Eq.~(\ref{vdot}) and Eq.~(\ref{cont2}) in Eq.~(\ref{teul}) to find
\begin{equation}
\partial_t \left(\frac{1}{\rho r^2} \partial_t f \right) + \partial_t \left(\frac{f}{\rho^2 r^4} \partial_r f\right) + \partial_r \left(\frac{f}{\rho^2 r^4} \partial_t f \right) + \partial_r \left(\left(\frac{f^2}{\rho^3 r^6} - \frac{c_s^2}{\rho r^2}\right)\partial_r f\right) =0 
\label{meq}
\end{equation}

If we now define the `inverse acoustic metric' components
\begin{equation}
g^{tt} = \frac{1}{\rho r^2} \,, \quad g^{rt} = \frac{f}{\rho^2 r^4} = g^{tr} \,, \quad g^{rr} = \frac{f^2}{\rho^3 r^6} - \frac{c_s^2}{\rho r^2} \,,
\label{m.comp}
\end{equation}

we find that Eq.~(\ref{meq}) has the expression
\begin{equation}
\partial_{\mu}\left(g^{\mu \nu} \partial_{\nu} f\right) = 0
\label{wav}
\end{equation}

Eq.~(\ref{wav}) is the wave equation that will feature in our perturbative approach. We note that the time derivative of the Euler equation does not generically lead to a wave equation as in Eq.~(\ref{wav}). The above equation is a consequence of spherically symmetric flows described in terms of the mass accretion rate $f$. 

On performing a perturbative expansion of the fields about a known solution, the inverse metric components in Eq.~(\ref{m.comp}) and the wave equation Eq.~(\ref{wav}) will shown to provide a compact representation for arbitrary order nonlinear perturbations of $f$ propagating on an effective dynamical analogue spacetime. We note here that the analogue spacetime should technically be associated with a metric $G_{\mu \nu}$, with the fluctuations satisfying a massless scalar field equation on the spacetime 

\begin{equation}
\frac{1}{\sqrt{-G}}\partial_{\mu}\left(\sqrt{-G}G^{\mu \nu} \partial_{\nu} f\right) = 0 \,.
\label{wave.2}
\end{equation}

The transformation of the metric components from Eq.~(\ref{wav}) to Eq.~(\ref{wave.2}) is well defined in more than two spacetime dimensions. We can use the spherical symmetry of the system to extend the dimensions of $g^{\mu \nu}$ in Eq.~(\ref{wav}) and find several possible definitions of $G^{\mu \nu}$ in Eq.~(\ref{wave.2}). For example, we could consider a four dimensional extension by introducing $g^{\theta \theta}$ and $g^{\phi \phi} = \frac{g^{\theta \theta}}{\sin \theta}$ components, which respects spherical symmetry and does not modify Eq.~(\ref{wav}). The transformation from Eq.~(\ref{wav}) to Eq.~(\ref{wave.2}) now follows from $G^{\mu\nu} = \sqrt{-g} g^{\mu \nu}$, with $g$ the determinant of the four dimensional metric. Regardless of the way in which the redundant dimensions are included, the components of $G^{\mu \nu}$ are always be proportional to $g^{\mu \nu}$ up to an overall factor that depends on the metric determinant. As this scaling neither affects the causal structure nor the underlying dynamical features, we prefer to consider the formalism with the unique two dimensional inverse metric $g^{\mu \nu}$ with components as in Eq.~(\ref{m.comp}).

Before proceeding to perturbatively solve Eq.~(\ref{cont2}) and Eq.~(\ref{wav}) about a stationary solution, we will discuss certain properties of the components in Eq.~(\ref{m.comp}) that are manifest throughout the flow. All solutions that we consider will have a non-negative mass accretion rate and density profile. This then implies that the components of $g^{tt}$ and $g^{tr}$ in Eq.~(\ref{m.comp}) never take on negative values. On the other hand, $g^{rr}$ can take on both positive and negative values, and we can determine that it vanishes when

\begin{equation}
r_H^2 = \frac{f}{\rho c_s} \,,
\label{horrad}
\end{equation}  

where $r_H$ denotes the acoustic horizon radius. From Eq.~(\ref{mac}), we see that $v = c_s$ at $r= r_H$ and hence the horizon is located at the critical point where the flow velocity matches the local sound speed of the fluid. $g^{rr}$ is negative in the subsonic region where $v<c_s$ and positive in the supersonic region where $v>c_s$. 

To solve Eq.~(\ref{cont2}) and Eq.~(\ref{wav}), we now consider the perturbative expansions of $f$ and $\rho$ about a known stationary solution up to $n^{\text{th}}$ order to be of the form
\begin{align}
f(r,t) &= f_0 + \sum_{k=1}^{n}\epsilon^k f_k(r,t) = f_0\left(1 + \frac{1}{f_0} \sum_{k=1}^{n}\epsilon^k f_k(r,t)\right) \label{fn} \\
\rho(r,t) & = \rho_0(r) + \sum_{k=1}^{n} \epsilon^k \rho_k(r,t) = \rho_0(r)\left( 1 + \frac{1}{\rho_0(r)}\sum_{k=1}^{n} \epsilon^k \rho_k(r,t)\right) \,, \label{rn}
\end{align}
where $f_0$ and $\rho_0$ denote the stationary solution values. In the stationary solution, $f_0$ is a constant while $\rho_0 = \rho_0(r)$ is independent of time. The dimensionless counting parameter $\epsilon$ measures the strength of the perturbation. The expansions in Eq.~(\ref{fn}) and Eq.~(\ref{rn}) determine those of all other dependent quantities --  such as the components of the inverse acoustic metric, the velocity, the speed of sound, etc. to any order in perturbation. If $A(r,t)$ is such a dependent field with the $n^{\text{th}}$ order expansion
\begin{equation}
A(r,t) = A_0(r) + \sum_{k=1}^{n}\epsilon^k A_k(r,t) = A_0\left(1 + \frac{1}{A_0(r)} \sum_{k=1}^{n}\epsilon^k A_k(r,t)\right) \,,
\end{equation}
then the expansion is perturbative provided
\begin{equation}
\epsilon \frac{\vert A_{l+1} \vert}{\vert A_{l} \vert} < 1 \,, \qquad  l = 0 \,, \cdots n-1
\label{pert.con}
\end{equation}

To determine the $n^{\text{th}}$ order solutions of $f$ and $\rho$, we first substitute the expansions of Eq.~(\ref{rn}) in Eq.~(\ref{cont2}) and Eq.~(\ref{wav}). We then collect the expressions in powers of $\epsilon$ and solve for the coefficients. The coefficient of $\epsilon^0$ is manifestly satisfied by the stationary solution. The coefficient of $\epsilon$ resulting from Eq.~(\ref{cont2}) is
\begin{equation}
\dot{\rho_1} + \frac{\partial_r f_1}{r^2} = 0 \,, \label{r1e}
\end{equation}
while from Eq.~(\ref{wav}) we find
\begin{equation}
\partial_{\mu}\left(g_{(0)}^{\mu \nu} \partial_{\nu} f_1\right) = 0 \,,
\label{wav.1}
\end{equation}
where 
\begin{equation}
g^{tt}_{(0)} = \frac{1}{r^2 \rho_0} \,, \qquad g^{tr}_{(0)} = \frac{f_0}{r^4 \rho_0^2} = g^{rt}_{(0)} \,, \qquad g^{rr}_{(0)} = \frac{f_0^2}{\rho_0^3 r^6}-\frac{c^2_{s0}}{ \rho_0 r^2}
\label{met0}
\end{equation}
The inverse metric $g^{\mu\nu}_{(0)}$ is completely determined by the stationary solution. The components also agree with the known stationary analogue spacetime on which the linear mass accretion rate fluctuation propagates. For reasons that will be shortly explained, we will consider Eq.~(\ref{met0}) as the inverse metric components of the zeroth order effective acoustic spacetime. We can solve Eq.~(\ref{wav.1}) for the unknown first order pertubation $f_1$, which is then used to determine $\rho_1$ from Eq.~(\ref{r1e}). In this way, the first order solutions can be consistently derived.

The solutions at higher orders follow iteratively.
The second order equations result from coefficients of $\epsilon^2$ on substituting the expansions in Eq.~(\ref{fn}) and Eq.~(\ref{rn}), which are 
\begin{equation}
\dot{\rho_2} + \frac{\partial_r f_2}{r^2} = 0 \,, \label{r2e}
\end{equation}
from Eq.~(\ref{cont2}) and
\begin{align}
\partial_{\mu}\left(g_{(0)}^{\mu \nu} \partial_{\nu} f_2\right) = - \partial_{\mu}\left(g_{(1)}^{\mu \nu} \partial_{\nu} f_1\right)
\label{f2e}
\end{align}
from Eq.~(\ref{wav}).  The coefficients of $g_{(1)}^{\mu \nu}$ are  
\begin{align}
g^{tt}_{(1)} = \frac{1}{r^2 \rho_0}\left(-\frac{\rho_1}{\rho_0}\right)& \,, \qquad g^{tr}_{(1)} = \frac{f_0}{r^4 \rho_0^2}\left(\frac{f_1}{f_0} - 2 \frac{\rho_1}{\rho_0}\right) = g^{rt}_{(1)} \,, \notag\\
g_{(1)}^{rr} =&\frac{f_0^2}{r^6 \rho_0^3} \left(2 \frac{f_1}{f_0} - 3 \frac{\rho_1}{\rho_0}\right)-\frac{c_{s0}^2}{\rho_0 r^2}\left(\frac{\rho_1}{c_{s0}^2}\frac{\partial c_{s}^2}{\partial \rho}\Bigg\vert_{\rho_0} - \frac{\rho_1}{\rho_0}\right)\,,\label{met1}
\end{align}
where $\frac{\partial c_{s}^2}{\partial \rho}\Big\vert_{\rho_0}$ represents the derivative of $c_s^2$ with respect to $\rho$, evaluated at $\rho_0$. At this order, we can once again solve Eq.~(\ref{f2e}) for $f_2$, following which we can derive $\rho_2$ from Eq.~(\ref{r2e}). The key difference with the first order perturbation equations concerns the presence of an effective source constructed from the first order solutions in Eq.~(\ref{f2e}). However Eq.~(\ref{wav}) allows us to interpret Eqs.~(\ref{wav.1}) and (\ref{f2e}) with no loss in generality in terms of a collective second order fluctuation $f_0 + \epsilon f_1 + \epsilon^2 f_2$ propagating on a first order effective acoustic spacetime with inverse metric $g_{(0)}^{\mu \nu} + \epsilon g_{(1)}^{\mu \nu}$. This interpretation provides a faithful representation of the perturbation equations up to this order, while allowing us to describe their solutions as a collective massless fluctuation propagating on an effective acoustic spacetime.  

The coefficients of $\epsilon^3$ in Eq.~(\ref{cont2}) and Eq.~(\ref{wav}) likewise provide us with the following third order equations
\begin{align}
\dot{\rho_3} + \frac{\partial_r f_3}{r^2} &= 0 \,, \label{r3e} \\
\partial_{\mu}\left(g_{(0)}^{\mu \nu} \partial_{\nu} f_3\right) &= - \partial_{\mu}\left(g_{(2)}^{\mu \nu} \partial_{\nu} f_1\right) - \partial_{\mu}\left(g_{(1)}^{\mu \nu} \partial_{\nu} f_2\right) \,,
\label{f3e}
\end{align}
with the inverse components of $g^{\mu \nu}_{(2)}$ given by
\begin{align}
g^{tt}_{(2)} &= \frac{1}{r^2 \rho_0}\left(\left(\frac{\rho_1}{\rho_0}\right)^2 -\frac{\rho_2}{\rho_0}\right) \,, \qquad  g_{(2)}^{tr} = \frac{f_0}{r^4 \rho_0^2}\left(\frac{f_2}{f_0} - 2 \left(\frac{\rho_2}{\rho_0} + \frac{f_1}{f_0}\frac{\rho_1}{\rho_0}\right) + 3 \left(\frac{\rho_1}{\rho_0}\right)^2\right) = g^{rt}_{(2)} \,,\notag\\
g^{rr}_{(2)} &= \frac{f_0^2}{r^6 \rho_0^3} \left(\left(\frac{f_1}{f_0}\right)^2 + 2 \frac{f_2}{f_0} - 3 \frac{\rho_2}{\rho_0} + 6\left(\left(\frac{\rho_1}{\rho_0}\right)^2 - \frac{f_1}{f_0}\frac{\rho_1}{\rho_0}\right)\right) \notag\\
&\qquad \qquad  - \frac{c_{s0}^2}{\rho_0 r^2}\left(\frac{\rho_2}{c_{s0}^2}\frac{\partial c_{s}^2}{\partial \rho}\Bigg\vert_{\rho_0} + \frac{\rho^2_1}{2 c_{s0}^2}\frac{\partial^2 c_{s}^2}{\partial \rho^2}\Bigg\vert_{\rho_0} - \frac{\rho_1}{c_{s0}^2} \frac{\rho_1}{\rho_0}\frac{\partial c_{s}^2}{\partial \rho}\Bigg\vert_{\rho_0} - \frac{\rho_2}{\rho_0} + \left(\frac{\rho_1}{\rho_0}\right)^2\right)
\label{met2}
\end{align}
The equations at this order can be solved for $f_3$ and $\rho_3$ as in the previous orders in pertubation. Eqs.~(\ref{f3e}), (\ref{f2e}) and (\ref{wav.1}) can be  described in terms of the collective fluctuation $f_0 + \epsilon f_1 + \epsilon^2 f_2 + \epsilon^3 f_3$ propagating on a second order effective acoustic spacetime with inverse metric $g_{(0)}^{\mu \nu} + \epsilon g_{(1)}^{\mu \nu} + \epsilon^2 g_{(2)}^{\mu \nu}$.

The iterative procedure continues to all orders. Assuming known solutions up to order $n-1$, we now collect the $\epsilon^n$ coefficient resulting from substituting Eq.~(\ref{fn}) and Eq.~(\ref{rn}) in Eq.~(\ref{cont2}) and Eq.~(\ref{wav}). This gives us the equations
\begin{align}
\dot{\rho_n} &= - \frac{\partial_r f_n}{r^2} \label{rne} \\
\partial_{\mu}\left(g^{\mu \nu}_{(0)} \partial_{\nu} f_n\right) &= - \sum_{k=1}^{n-1} \partial_{\mu}\left(g^{\mu \nu}_{(k)} \partial_{\nu} f_{n-k}\right) \label{fne}
\end{align}
The total mass accretion rate $f$ up to order $n$ propagates on an effective acoustic spacetime with inverse metric $g_{\text{eff}(n-1)}^{\mu\nu}$ constructed from the preceding $n-1$ orders in perturbation
\begin{equation}
g_{\text{eff}(n-1)}^{\mu\nu} = \sum_{k=0}^{n-1} \epsilon^k g^{\mu\nu}_{(k)}
\label{nm1.met}
\end{equation}

\section{Properties of the dynamical acoustic spacetime} \label{curve}

In the previous section, we determined that an effective acoustic spacetime can be described to every order following nonlinear perturbations about a stationary flow. The properties of this dynamical spacetime to all orders is most conveniently investigated using Eq.~(\ref{m.comp}), with the understanding that the components at a particular order result from using the expansions in Eq.~(\ref{fn}) and Eq.~(\ref{rn}). In this section, we will consider the causal structure and curvature of the effective acoustic spacetime, and the fluctuation relation satisfied by its acoustic horizon.

From Eq.~(\ref{m.comp}), we find the following components of the effective acoustic spacetime metric
\begin{align}
g_{\mu\nu} &= 
\begin{pmatrix}
g_{tt} & g_{tr} \\
g_{rt} & g_{rr}
\end{pmatrix}
=
\begin{pmatrix}
 g g^{rr} & - g g^{rt} \\
- g g^{tr} & g g^{tt}
\end{pmatrix} \notag\\
&=
\begin{pmatrix}
\rho r^2 \left(1 - \beta^2\right) & \frac{f}{c_s^2}\\
\frac{f}{c_s^2} & -\frac{\rho r^2}{c_s^2}
\end{pmatrix} \,,
\label{met.mat}
\end{align}
where $g = - \frac{\rho^2 r^4}{c_s^2}$ is the determinant of the metric and $\beta = \frac{f}{c_s \rho r^2}$ is the ratio of the fluid velocity to the sound speed (recalling from Eq.~(\ref{mac}) that $v = \frac{f}{\rho r^2}$). Since the fluid velocity can range from $0<v<1$, in units where the speed of light $c=1$ , we have $0< \beta < \frac{1}{c_s}$.

We define the line element in the usual way
\begin{equation}
ds^2 = g_{\mu \nu} dx^{\mu} dx^{\nu} = g_{tt}dt^2 + 2 g_{tr} dt dr + g_{rr}dr^2
\label{line}
\end{equation}

The causal structure of this metric can be made manifest through its expression in terms of ingoing and outgoing null directions, which can be determined from considering $ds^2 = 0$ in Eq.~(\ref{line}). The quadratic equation in $dt$ admit two roots
\begin{equation}
dt = \left(\frac{g_{tr}}{g_{tt}} \pm \sqrt{\left(\frac{g_{rt}}{g_{tt}}\right)^2 - \frac{g_{rr}}{g_{tt}}}\right) dr
\end{equation}
Using the metric components in Eq.~(\ref{met.mat}) we find the solutions
\begin{align}
dt &= \frac{1}{c_s+ \frac{f}{\rho r^2}}dr \,, \label{dwn}\\
dt &= -\frac{1}{c_s - \frac{f}{\rho r^2}} dr \label{up}
\end{align}

Hence Eq.~(\ref{dwn}) and Eq.~(\ref{up}) describe motion with and against the fluid respectively, which characterizes the causal properties of this spacetime with respect to the flow. These solutions provide the definitions of ingoing and outgoing null coordinates
\begin{equation}
d\bar{u} = \frac{1}{\sqrt{2}}\left(dt - \frac{1}{c_s + v}dr\right) \,, \qquad d\bar{v} = \frac{1}{\sqrt{2}}\left(dt + \frac{1}{c_s - v}dr\right) \,,
\label{ioc}
\end{equation}

with which the sonic line element in Eq.~(\ref{line}) can be expressed as
\begin{equation}
ds^2 = 2 d\bar{u} d\bar{v}
\end{equation}

It can be noted from Eqs.~(\ref{dwn}) and (\ref{up}), that while both ingoing and outgoing flows can exist in the subsonic region $v<c_s$, we can only have ingoing flows in the supersonic region $v>c_s$.   

We can also demonstrate that there exists a non-trivial curvature of the spacetime. A 2 dimensional Riemmanian spacetime has only one independent curvature term - the Ricci scalar $R$. The Riemann tensor is expressed in terms of $R$ as
\begin{equation}
R_{t r t r} = \frac{1}{2}\left( g_{tt} g_{rr} - g_{tr}^2\right) R
\end{equation}

To derive the curvature, we will need to determine the connection components
\begin{equation}
\Gamma^{\alpha}_{\mu \nu} = \frac{1}{2}g^{\alpha \beta} \left(g_{\mu \beta\,, \nu} + g_{\nu \beta\,, \mu} - g_{\mu \nu\,, \beta}\right) \label{conn}
\end{equation}

and subsequently the Ricci tensor components and Ricci scalar
\begin{align}
R_{\mu\nu} &= \Gamma^{\alpha}_{\mu\nu\,, \alpha} - \Gamma^{\alpha}_{\mu \alpha\,, \nu} + \Gamma^{\alpha}_{\beta \alpha}\Gamma^{\beta}_{\mu\nu} - \Gamma^{\alpha}_{\beta \nu}\Gamma^{\beta}_{\mu \alpha} \,, \label{r.t}\\
R &= g^{\mu\nu}R_{\mu\nu} = g^{tt}R_{tt} + 2g^{tr}R_{tr} + g^{rr}R_{rr} \,. \label{r.s}
\end{align}

The connection components can be computed to be
\begin{align}
\Gamma^t_{tt} = \frac{1}{2}\frac{g_{rr}\dot{g}_{tt}+ g_{tr}\left(g'_{tt} - 2 \dot{g}_{tr}\right)}{g} &\,, \qquad \Gamma^t_{tr} = \frac{1}{2}\frac{g_{rr}\dot{g}'_{tt} - g_{tr}g'_{rr}}{g} \notag\\
\Gamma^t_{rr} = -\frac{1}{2}\frac{g_{tr}g'_{rr} + g_{rr}\left(\dot{g}_{rr} - 2 g'_{tr}\right)}{g} &\,, \qquad \Gamma^r_{tt} = -\frac{1}{2}\frac{g_{tr}\dot{g}_{tt}+ g_{tt}\left(g'_{tt} - 2 \dot{g}_{tr}\right)}{g}\notag\\
\Gamma^r_{rt} = -\frac{1}{2}\frac{g_{tr}g'_{tt} - g_{tt}\dot{g}_{rr}}{g} &\,,\qquad \Gamma^r_{rr} = -\frac{1}{2}\frac{g_{tt}g'_{rr}+ g_{tr}\left(\dot{g}_{rr} - 2 g'_{tr}\right)}{g}
\label{conn.comp}
\end{align}

Using Eq.~(\ref{r.t}) and Eq.~(\ref{r.s}) we hence find the following Ricci scalar
\begin{align}
R & = \frac{1}{2\left(g_{tt}g_{rr} - g_{rt}\right)^2}\Bigg[g_{tr}\left(g'_{rr}\dot{g}_{tt} - 2g_{tr}\left(g'_{tt} - 2 \dot{g}_{tr}\right) - \dot{g}_{rr}\left(g'_{tt} + 2 \dot{g}_{tr}\right)\right) + g_{tt}\left(g'_{rr}\left(g'_{tt} - 2 \dot{g}_{tr}\right) + \dot{g}^2_{rr}\right)\notag\\
&\qquad \qquad + 2 g^2_{tr}\left(g''_{tt} - 2 \dot{g}'_{tr} + \dot{g}_{rr}\right) + g_{rr}\left(g'^2_{tt} + \dot{g}_{tt}\left(\dot{g}_{rr} - 2g'_{tr}\right) -2 g_{tt}\left(g''_{tt} - 2 \dot{g}'_{tr} + \ddot{g}_{rr}\right)\right)\Bigg]
\label{R.scalar}
\end{align} 

The zeroth order effective acoustic spacetime has a time independent curvature on account of being constructed from the stationary flow solution. However, the effective acoustic spacetime resulting from second and higher order perturbations of the fluid has a time dependent curvature. From Eq.~(\ref{R.scalar}), we see that the curvature remains well defined since the overall factor not in the parenthesis involves the square of the metric determinant, which is finite across the flow to all perturbation orders.
We also note here that spacetime metrics resulting from gravitational theories are subject to dynamical equations and constraints, which lead to the known result that they admit no dynamical degrees of freedom in $2$ dimensions. Acoustic metrics on the other hand are not subject to similar constraints hence have non-trivial dynamics in this case.

We conclude this section with an expression for the fluctuation of the acoustic horizon. On varying the horizon radius in Eq.~(\ref{horrad}), we find the following expression for the relative fluctuation of the acoustic horizon
\begin{equation}
\frac{\delta r_H}{r_H} = \frac{1}{2}\frac{\delta f}{f} - \frac{1}{4}\frac{\delta \rho}{\rho} \left(\frac{1}{c_s^2} \frac{\partial c_s^2}{\partial \rho} \Bigg \vert_{\rho} + 2\right)\,.
\label{varhor}
\end{equation}
The content of this equation is as follows. Given a known radial distance of the horizon at any order in perturbation, which satisfies Eq.~(\ref{horrad}) when expanded to that order, we can determine the change in the horizon radius relative to another order in perturbation from Eq.~(\ref{varhor}). Furthermore, Eq.~(\ref{varhor}) is only sensitive to the relative change in going between the perturbation orders. In the following, we assume that this relative change is considered from a given order in perturbation to the next order in perturbation. 

Using Eq.~(\ref{cont2}), we deduce 
\begin{equation}
\frac{\delta f}{f} = \frac{\int dr r^2 \delta \dot{\rho}}{\int dr r^2 \dot{\rho}}\,.
\label{fcont}
\end{equation}
The relative change in the mass accretion $\frac{\delta f}{f}$ is thus related to a spatially averaged change in energy flux and is expected to be positive in going to the next order in perturbation. This can be made more precise by considering a fluid that flows from an outer boundary to an inner boundary, as in the case of an accreting fluid. In this case, the averaged energy flux will be positive towards the inner boundary leading to a positive contribution to the change in the horizon radius. The relative change in the density $\frac{\delta \rho}{\rho}$ however is not similarly constrained. A given change in energy between perturbation orders could manifest either through an increase in the fluid velocity or density. We typically expect the velocity to increase towards the inner boundary in an accreting fluid, which would cause the change in density to decrease in this case. However, there could arise cases in which the relative change in density is positive and greater than $\frac{\delta f}{f}$.
Eq.~(\ref{varhor}) in this case predicts a receding acoustic horizon under perturbations. In the following section, we will demonstrate that certain low frequency nonlinear perturbations of the Bondi transonic flow solution can cause the density fluctuation to increase behind the horizon, thereby causing the acoustic horizon to recede. 

\section{Numerical solutions} \label{num}
In this section, we will numerically investigate solutions of $f$, $\rho$ and the inverse effective acoustic metric for the case of exponentially damped in time perturbations of the accreting Bondi flow solution. The unperturbed flow undergoes spherically symmetric accretion due to the presence of a central gravitating source. The accretor has a Newtonian potential, $\Phi(r) = -\frac{GM}{c^2 r}$, where $M$ denotes the mass of the object, $G$ is Newton's constant, $c$ is the speed of light and the radial distance $r$ now represents the distance from the center of the accretor placed at the origin. We normalize the length scale associated with the accretor, $r_g = \frac{GM}{c^2}$, by working in units where $G = M = c =1$. 

For accretion flows, we typically consider the equation of state to be either isothermal or isentropic. In the following, we assume that the flow is adiabatic and hence satisfies the isentropic condition $P= \kappa \rho^{\gamma}$ \footnote{The adiabatic exponent $\gamma$ is the ratio of the constant specific heats $c_P$ and $c_V$ of the fluid, while $\kappa$ is a measure of the fluid's constant specific entropy. For the Bondi flow, there is in addition the allowed range $\frac{4}{3} \le \gamma \le \frac{5}{3}$.}. With the isentropic condition, the local adiabatic sound speed of the fluid in Eq.~(\ref{ss}) takes the specific form
\begin{equation}
c_s^2 = \frac{\partial P}{\partial \rho} = \kappa \gamma \rho^{\gamma - 1}
\label{ss.2}
\end{equation}

The Bondi solution can be determined by specifying three constants, $\gamma$, $\kappa$ and the Bernoulli constant 
\begin{equation}
E = \frac{v^2}{2} + \frac{\kappa \gamma}{\gamma-1}\rho^{\gamma-1} + \Phi(r)
\label{en}
\end{equation}

In the following, we set $\gamma=1.35$, $\kappa=1$ and $E = 1.001$ throughout the flow. The solutions are then determined about the critical point where $v = c_s$. The inner boundary is $r=1$ and we take the maximum radius of $R_{\infty}=100$ to denote the outer boundary of the accreting fluid. With this setup, we determine the $v_0(r)$ and $c_{s0}(r)$ solutions whose plots are given in Figure 1.

\begin{figure}
\begin{center}
\includegraphics[width=0.43\columnwidth]{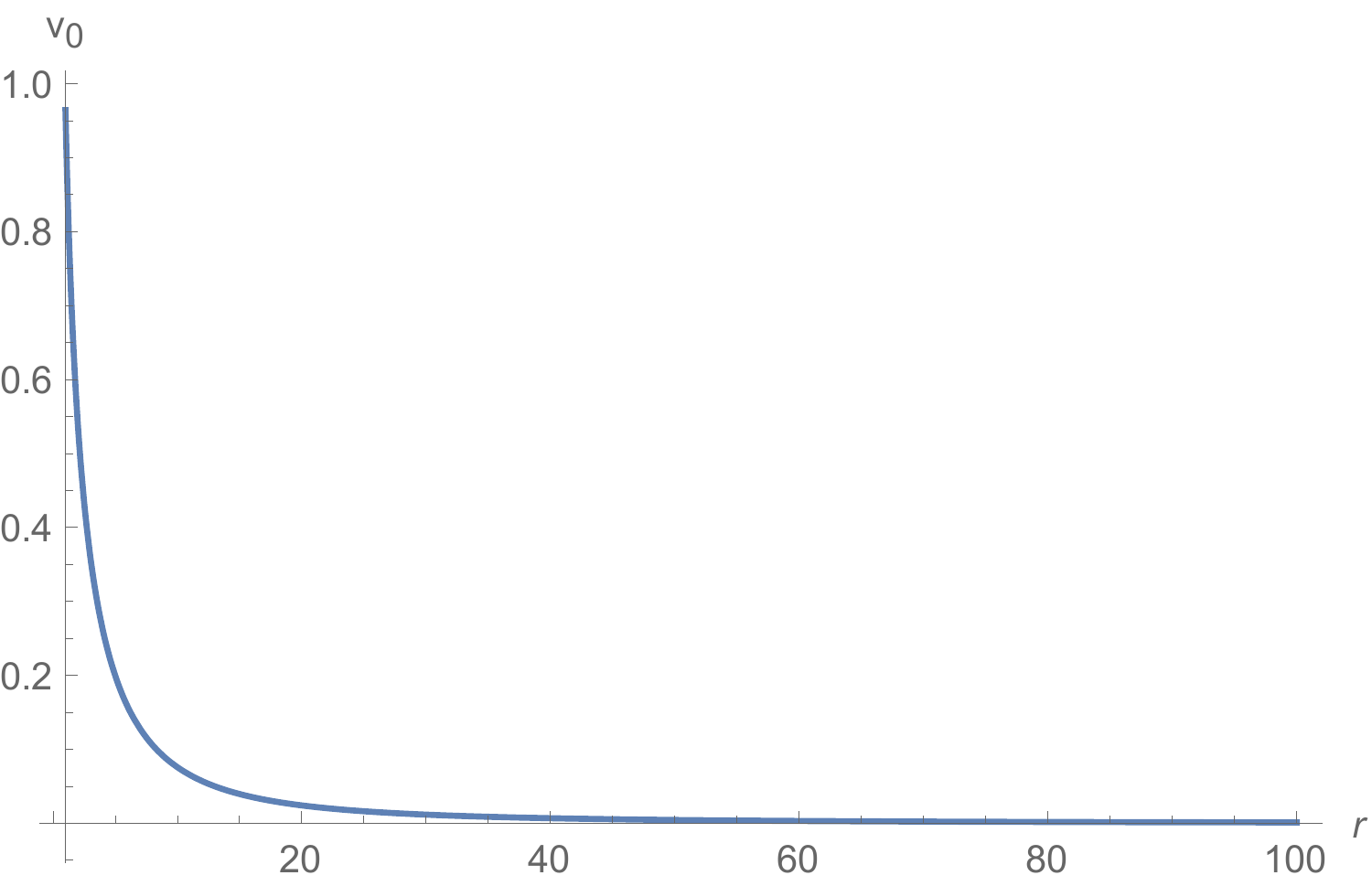}
\includegraphics[width=0.43\columnwidth]{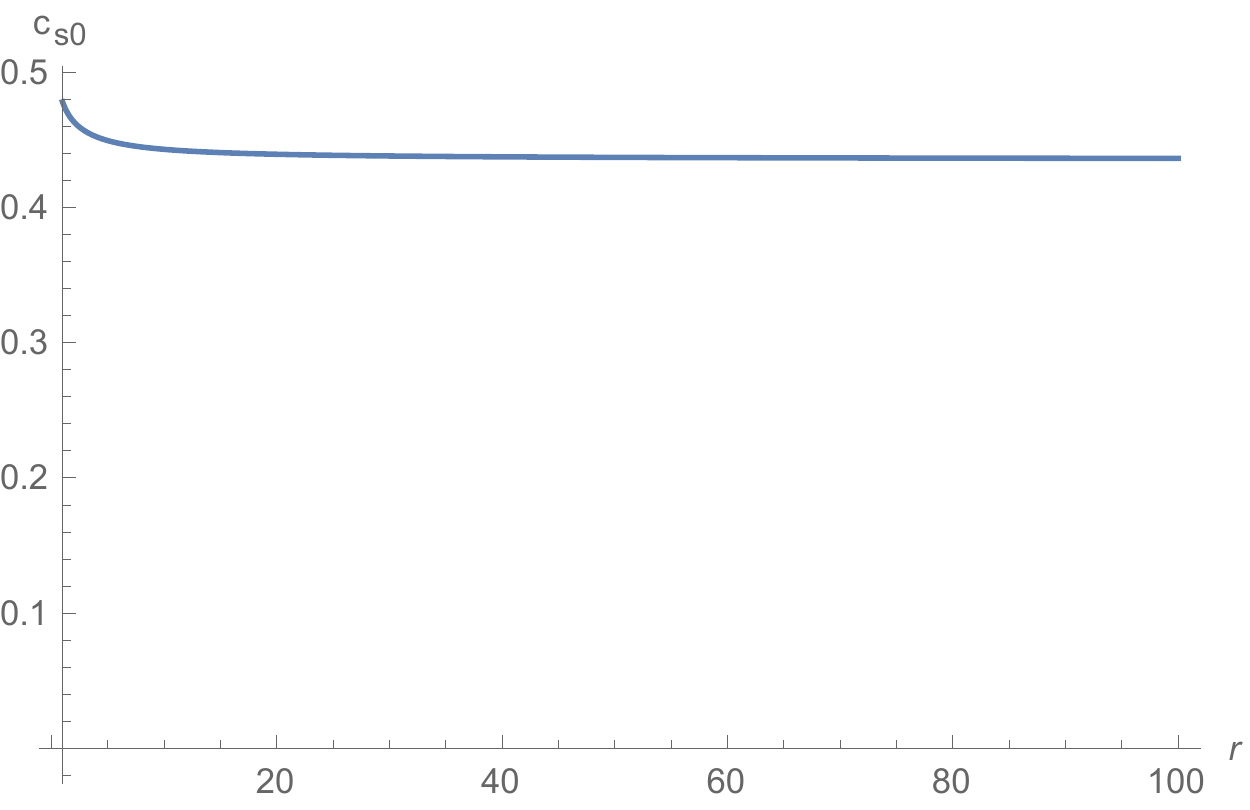}
\caption{Plots for $v_0(r)$ (left) and $c_{s0 }(r)$ (right). $(1,0)$ is the origin for both plots.}
\end{center}
\end{figure}

From these solutions for the fluid velocity and sound speed, we can uniquely determine the expressions of $f_0$ and $\rho_0$ using Eq.~(\ref{mac}) and Eq.~(\ref{ss}). We find that $f_0 = 0.0129$ and the solution for $\rho_0(r)$ is given in Figure 2.

\begin{figure}
\begin{center}
\includegraphics[width=0.89\columnwidth]{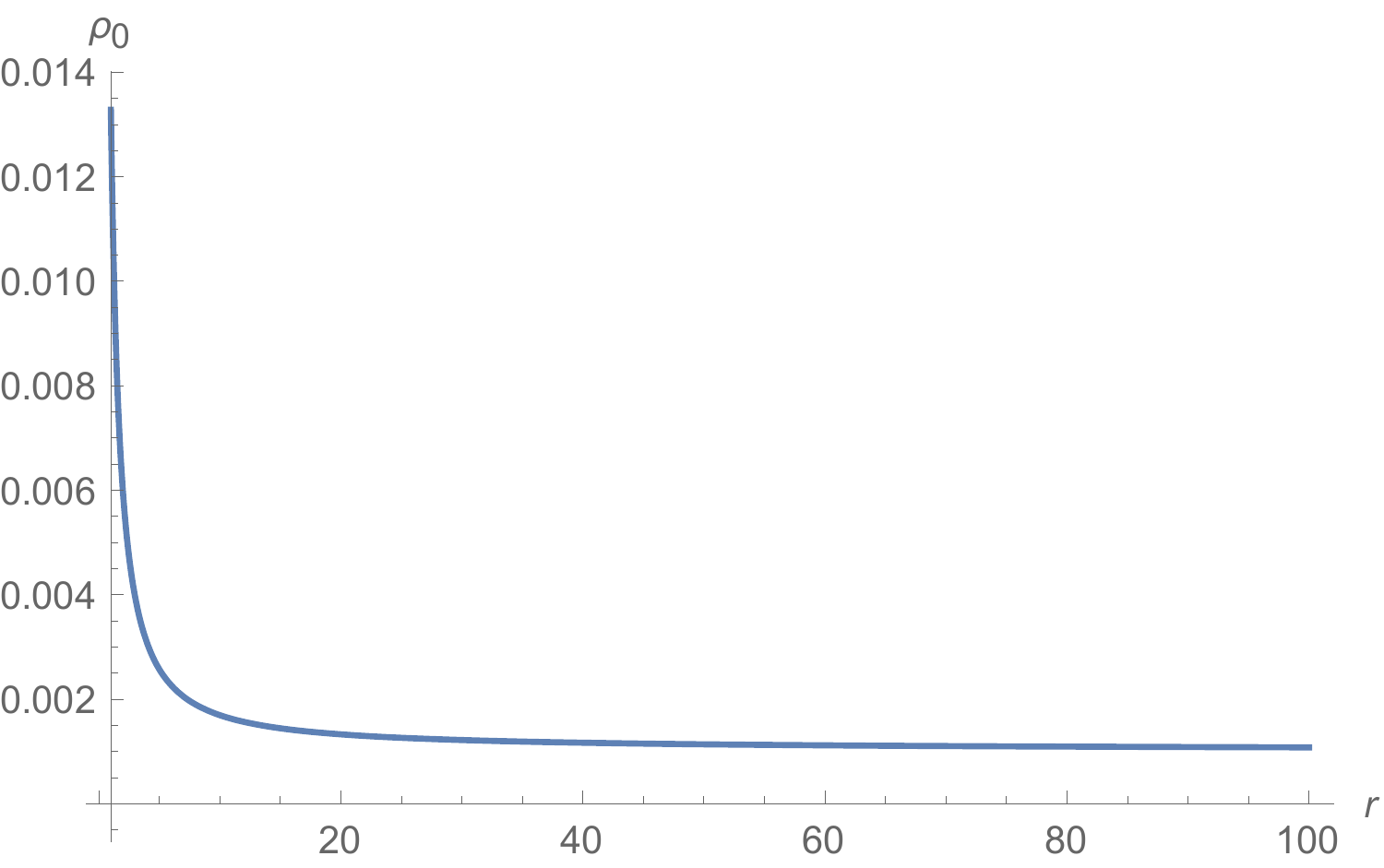}
\caption{Plot of $\rho_0(r)$. The origin is at $(1,0)$.}
\end{center}
\end{figure}

The transonic solution also determines the lowest order metric components using Eq.~(\ref{met0}). This is the zeroth order effective acoustic metric on which first order mass accretion rate perturbations propagate. At this order, the inverse metric components are time independent with solutions as given in Figure 3. We find that the lowest order acoustic horizon $g^{rr}_{(0)} = 0$ is located at $r_0=2.362$. It can also be noted that $g^{rr}_{(0)}>0$ for $r<r_0$ and $g^{rr}_{(0)}<0$ for $r>r_0$. However, $g^{tt}_{(0)}$ and $g^{tr}_{(0)}$ remain positive definite throughout the flow.

\begin{figure}
\begin{center}
\includegraphics[width=0.63\columnwidth]{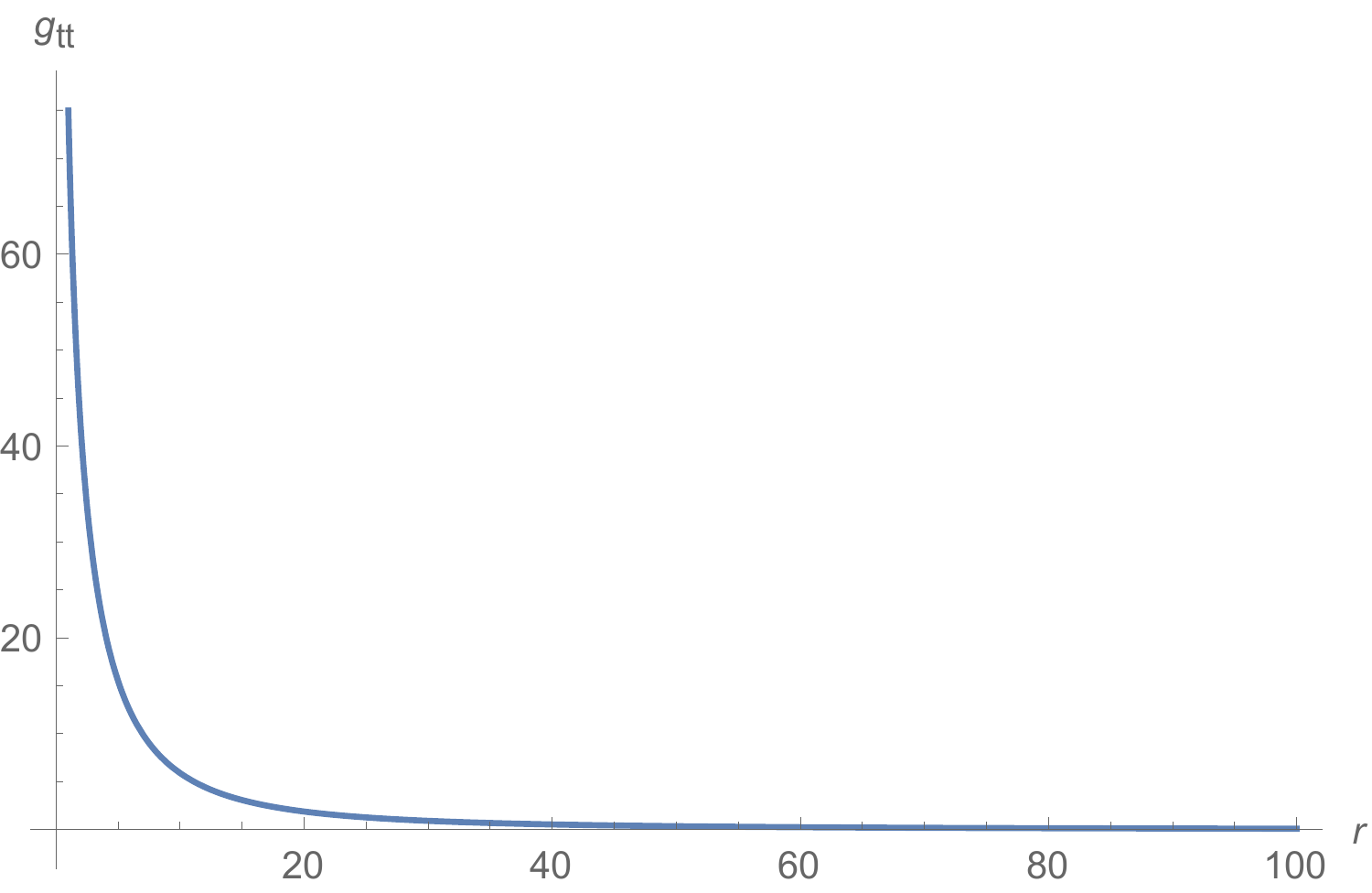}
\includegraphics[width=0.63\columnwidth]{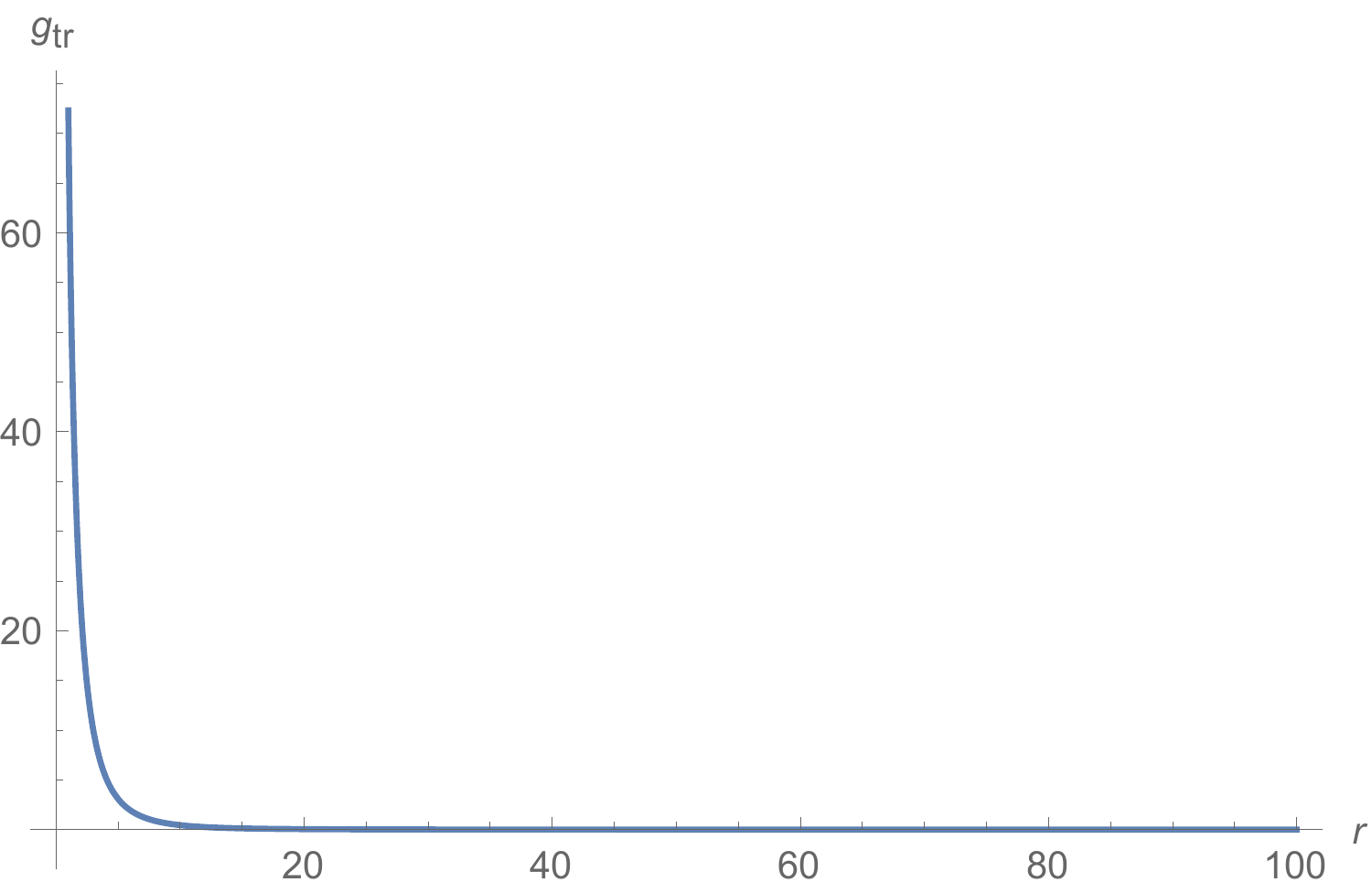}
\includegraphics[width=0.63\columnwidth]{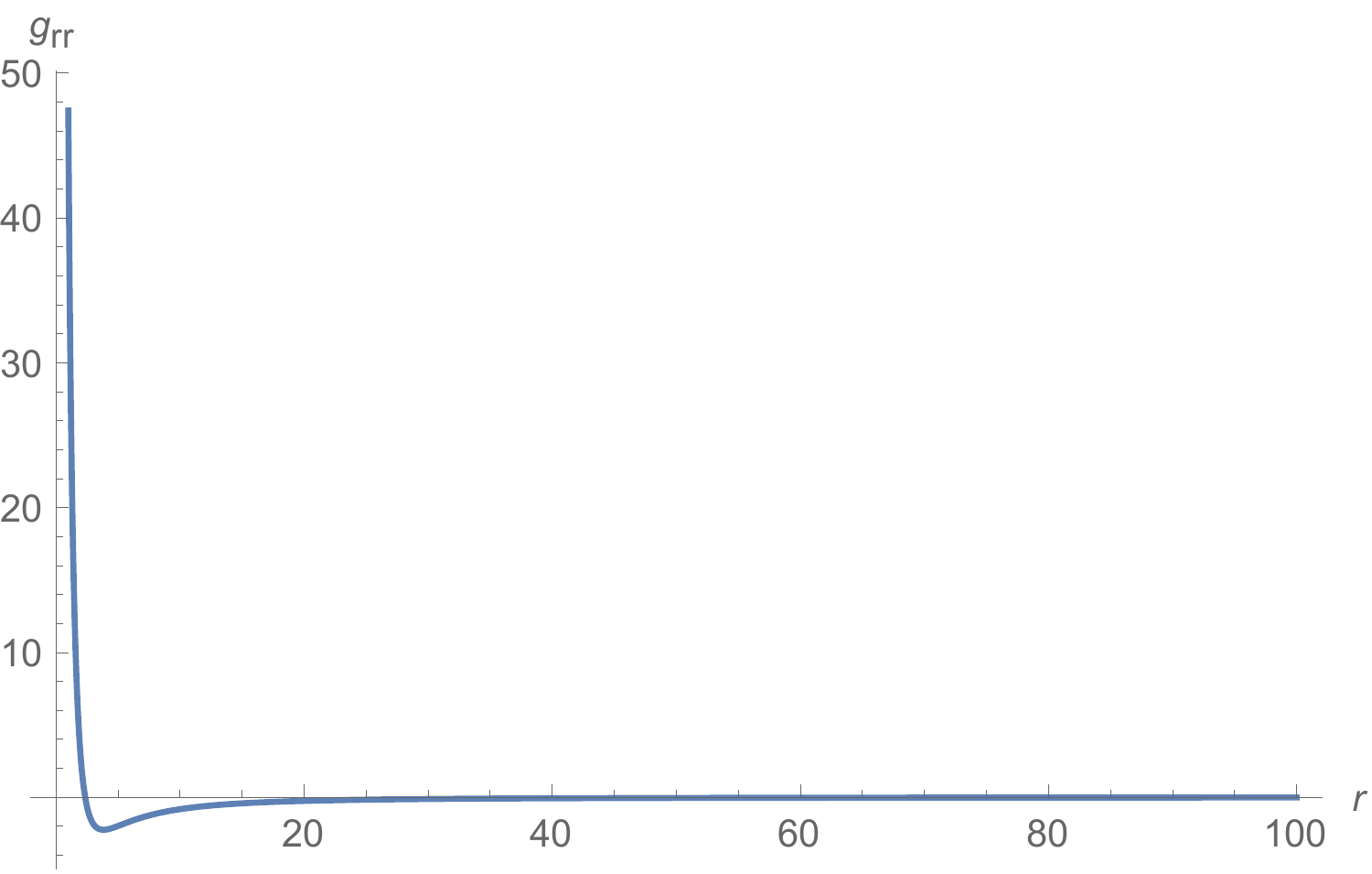}
\caption{Plots of inverse metric components $g^{tt}_{(0)}(r)$ (upper), $g^{tr}_{(0)}(r)$ (middle) and $g^{rr}_{(0)}(r)$ (lower). The origin is at $(1,0)$ in all plots. The location where $g^{rr}_{0} = 0$ is at $r=2.362$.}
\end{center}
\end{figure}

The zeroth order effective acoustic spacetime is a result of perturbations of the Bondi flow accreting solution, which we will now describe. The perturbations are considered exponentially damped in time $e^{-\omega t}$, which in particular represent a class of perturbations that hold for large times. 
The boundary conditions at $t=0$ are chosen to agree with the preceding order of perturbation. We thus impose $f_{l+1}(r,0) = f_l(r,0)$ and $\rho_{l+1}(r,0) = \rho_l(r,0)$ for $l=1\,,\cdots n$.
The perturbation is introduced at $t=0$ and considererd up to $t=10^3$, an order of magnitude larger than the spatial range, which is taken to our definition of `late time'. This choice ensures that the damping is neither inconsequential, nor overwhelms the background solution. The inner and outer spatial boundaries of the accreting fluid further allow us to consider two frequency ranges. `High frequency' perturbations $\omega_{\text{high}} \ge 1$ involve wavelengths that are the size of the accretor boundary or lower, while `Low frequency' perturbations $\omega_{\text{low}} \le 10^{-2}$ involve wavelengths larger that the radius of the accreting fluid. We will set $\omega_{\text{high}} = 10^{3}$ and $\omega_{\text{low}} = 10^{-3}$. As will be seen, in both frequency regimes the first order inverse metric solution for $g^{rr}_{(1)}$ has the largest amplitude at a particular instant of time $t_m = 632$ and $r = 1$. We use this maximum value in Eq.~(\ref{pert.con}) to fix $\epsilon$. In the following, we choose this strength in both frequency regimes to be $0.3$ and thus we accordingly have
\begin{equation}
 \epsilon \frac{\vert g^{rr}_{(1)}\vert_{(r=1, t=t_m)}}{\vert g^{rr}_{(0)}\vert_{(r=1)}} := 0.3
\label{npert}
\end{equation}
This choice further represents our consideration of `strong perturbations'. In taking the strength to be closer to $1$, the order of magnitude of $\epsilon$ in either frequency regime is the same as that resulting from Eq.~(\ref{npert}). Had we considered weak perturbations, then it would readily follow that the first order effective acoustic spacetime provides a good approximation to the dynamical acoustic spacetime at higher orders. In the following cases, we demonstrate that the near acoustic horizon dynamics of the first order effective acoustic spacetime remains a good approximation, with no significant corrections resulting from higher orders in perturbation. As a result, the main result in the low frequency analysis of a receding horizon is not affected by changing the strength of the perturbation.

\subsection{High frequency perturbations} \label{hfp}
We will set $\omega = \omega_{\text{high}} = 10^3$ throughout the high frequency analysis with $t \in \{0, 10^3\}$. In solving Eq.~(\ref{wav.1}), we set the boundary conditions $f_1(r,0) = f_0$ and choose $e^{-\omega t}$ at initial time from $r=1$ to $r=100$. This results in the solution plotted in Figure 4. In the subsonic region ($r>2.362$), we see the interference of growing and decaying modes with an increasing amplitude away from the horizon. This observation has been noted previously in travelling wave solutions and is a generic property of subsonic mass accretion rate fluctuations~\cite{Petterson:1980}. However in the supersonic region, only the ingoing modes propagate and the averaged mass accretion rate fluctuation increases towards the accretor. This observation agrees with our expectation from the fluctuation relation given in Eq.~(\ref{fcont}).

We can now use this result for $f_1$ to derive the solution for $\rho_1$ from the continuity equation in Eq.~(\ref{r1e}). This solution is plotted in Figure 5. The mass density fluctuation $\rho_1$ is relevant only in the supersonic flow, which can be anticipated on the basis of the contructive interference of $f_1$ in this region. We also note that $\rho_1$ is negative close to the accretor boundary. As the perturbations have wavelengths far smaller than the accretor boundary, they can propagate into the accretor thereby causing the mass density to decrease as in the solution. 

\begin{figure}[H]
\begin{center}
\includegraphics[width=0.8\columnwidth]{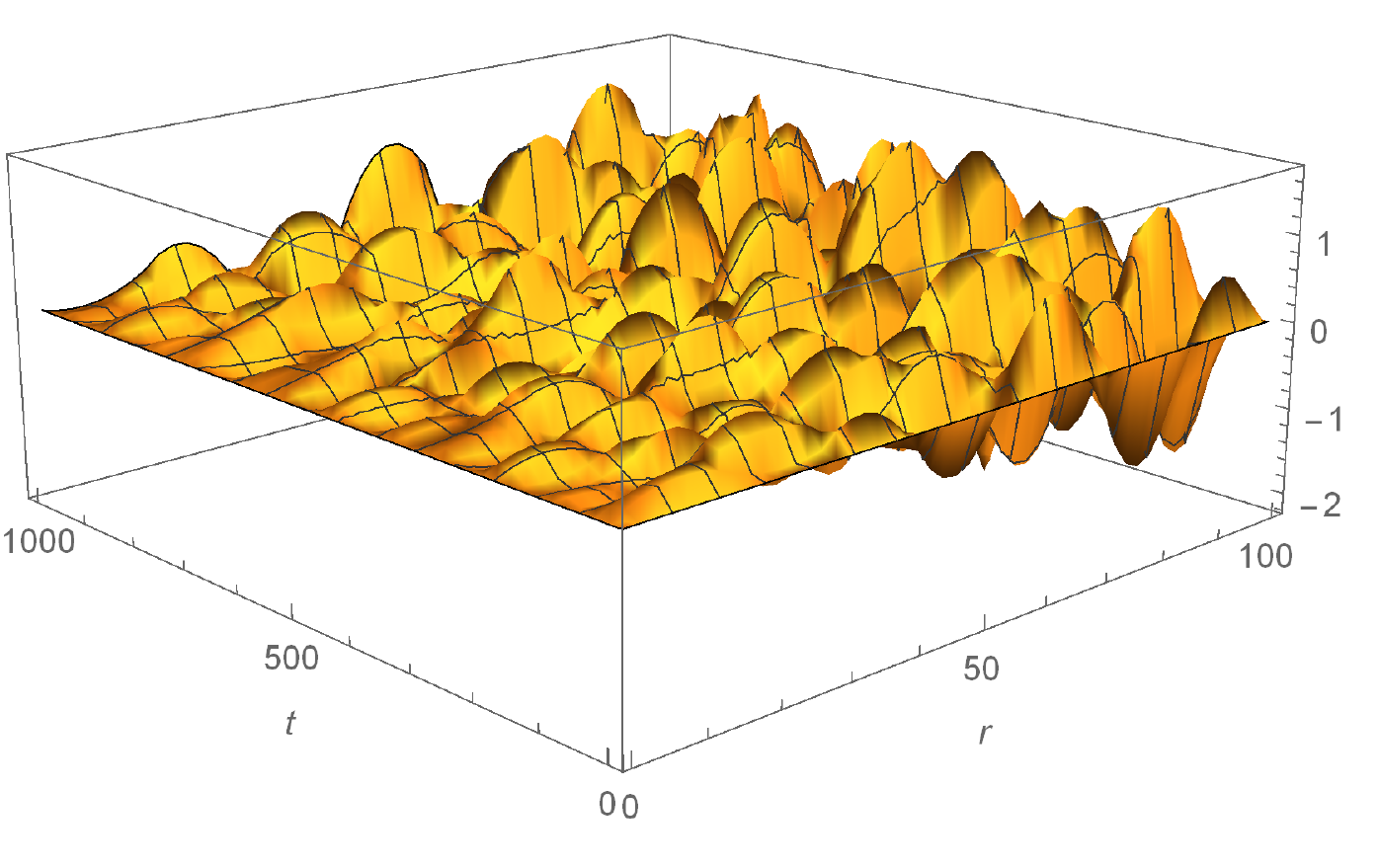}
\caption{Solution of the first order pertubed mass accretion rate $f_1(r,t)$.}
\end{center}
\end{figure}
\begin{figure}
\begin{center}
\includegraphics[width=0.8\columnwidth]{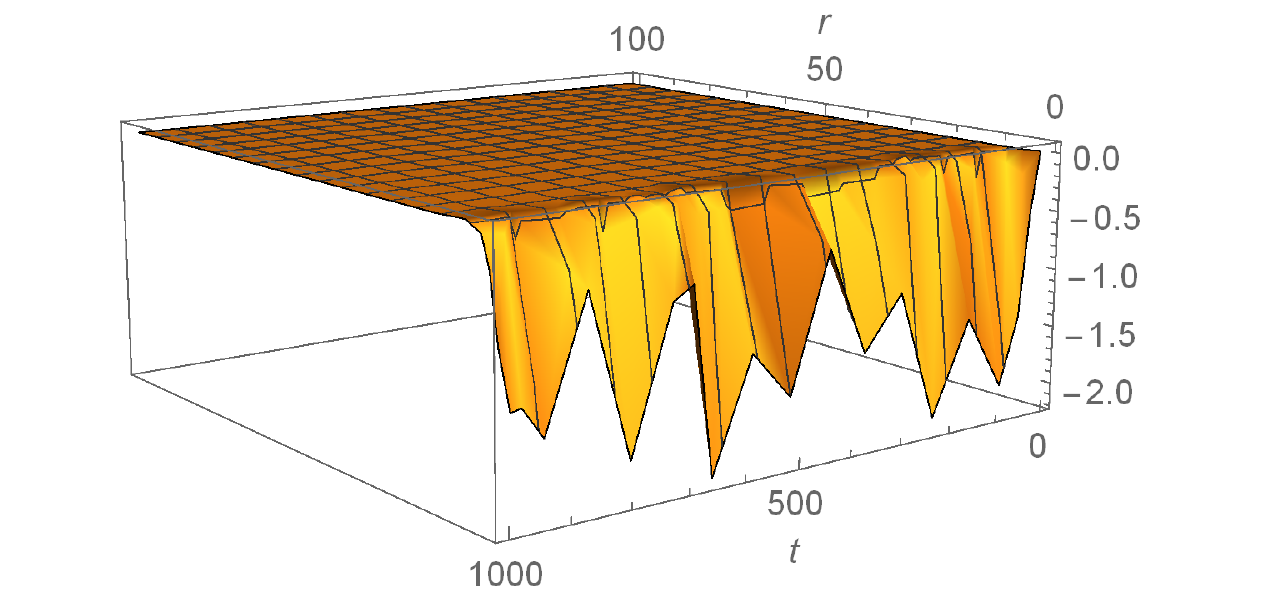}
\caption{Solution of the first order perturbed density $\rho_1(r,t)$. The perturbation is most relevant near the accretor boundary.}
\end{center}
\end{figure}
The decreasing density alone implies a reduced sound speed, while in conjuction with an increasing mass accretion rate it implies an increasing fluid velocity. This observation and the relative fluctuation equation in Eq.(\ref{varhor}), thus lead us to expect a first order effective acoustic horizon larger than the zeroth order solution, with a time dependent profile.

To explore this, we substitute the stationary and first order perturbation solutions of $f$ and $\rho$ in the inverse metric component expressions given in Eq.~(\ref{met1}). These components are plotted in Figure 6. The largest magnitude is for $g^{rr}_{(1)}$ at $r=1$ and $t = t_m = 632$. We fix $\epsilon$ for the entire perturbation scheme following Eq.~(\ref{npert}) to find $\epsilon \sim 4 \times 10^{-4}$.

\begin{figure}
\begin{center}
\includegraphics[width=0.8\columnwidth]{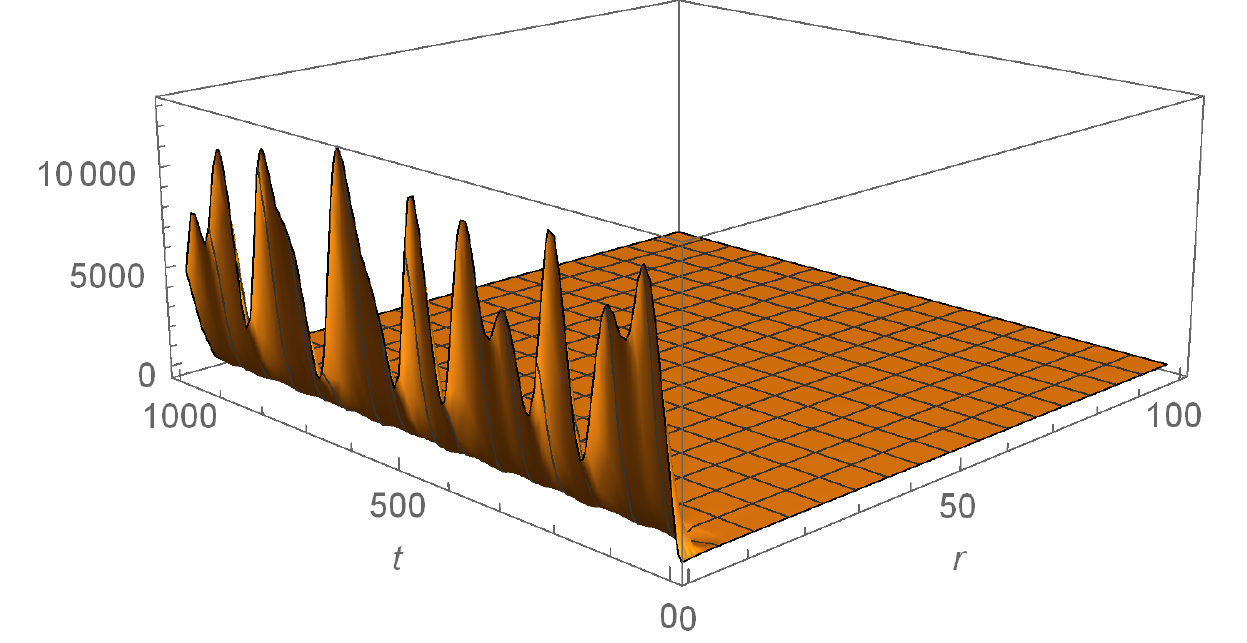}
\includegraphics[width=0.8\columnwidth]{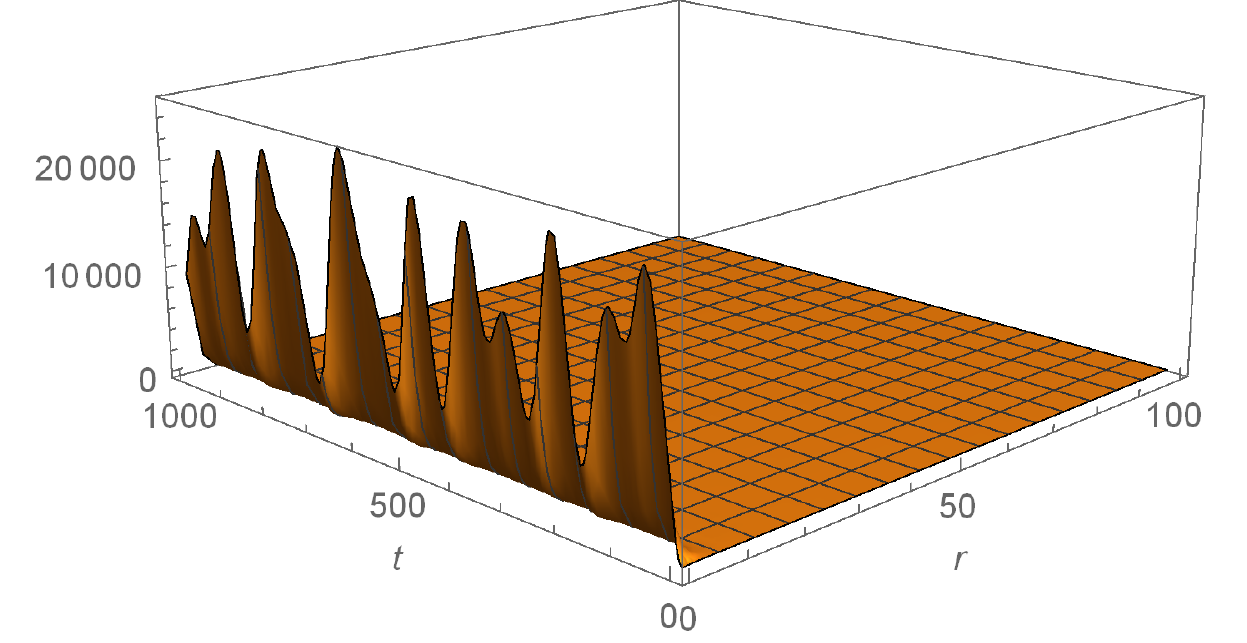}
\includegraphics[width=0.8\columnwidth]{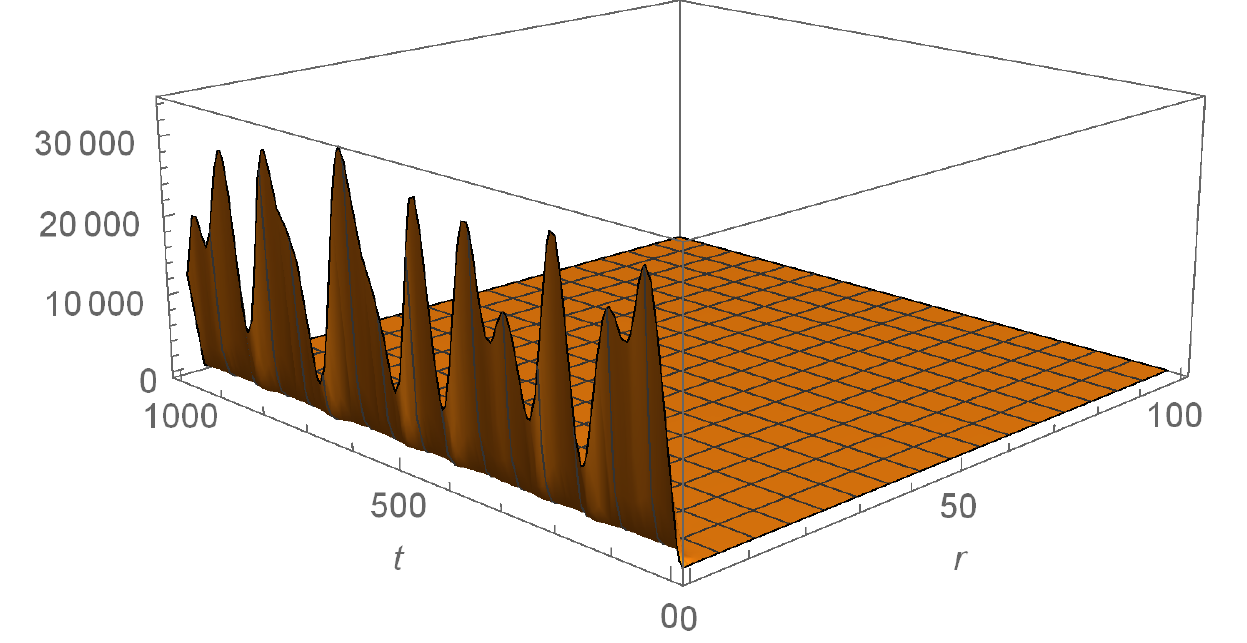}
\caption{First order perturbed inverse metric components : $g^{tt}_{(1)}(r,t)$ (upper), $g^{tr}_{(1)}(r,t)$ (middle) and $g^{rr}_{(1)}(r,t)$ (lower).}
\end{center}
\end{figure}

Using Eq.~(\ref{nm1.met}) and the metric components in Eq.~(\ref{met1}), we can construct the first order effective acoustic metric on which the collective second order mass accretion fluctuation propagates
\begin{equation}
g_{\text{eff}(1)}^{\mu\nu} =  g^{\mu\nu}_{(0)} +  \epsilon g^{\mu\nu}_{(1)} \,.
\label{nm1.met1}
\end{equation} 
The corresponding plots of the first order effective acoustic metric Eq.~(\ref{nm1.met1}) close to the acoustic horizon are given in Figure 7. We see that there are small time dependent perturbations that are only relevant in the supersonic region. While the $g_{\text{eff}(1)}^{tt}$ and $g_{\text{eff}(1)}^{tr}$ components are time dependent, they remain positive definite throughout. The acoustic horizon $g_{\text{eff}(1)}^{rr} = 0$ fluctuates about its stationary value.

\begin{figure}
\begin{center}
\includegraphics[width=0.6\columnwidth]{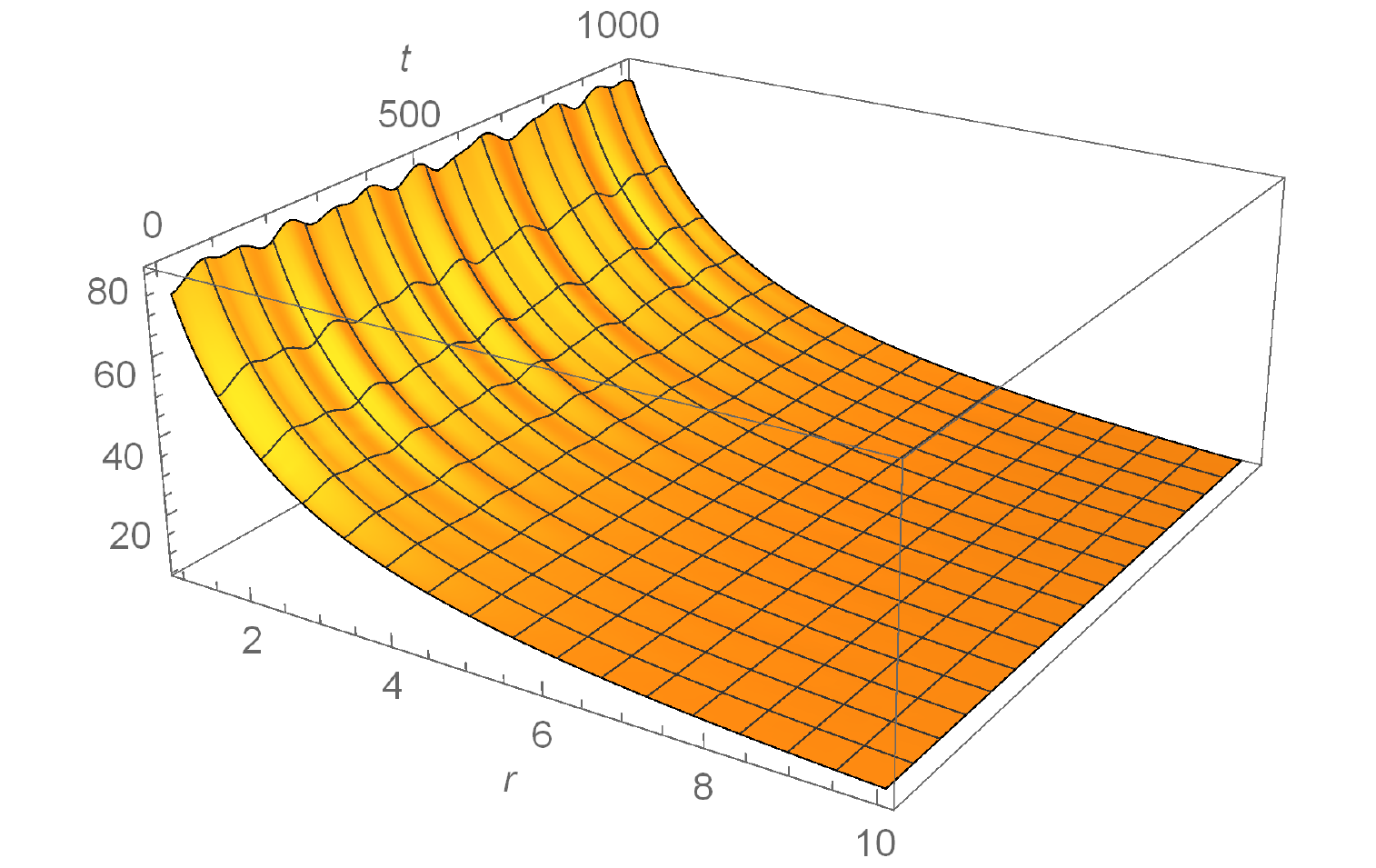}
\includegraphics[width=0.6\columnwidth]{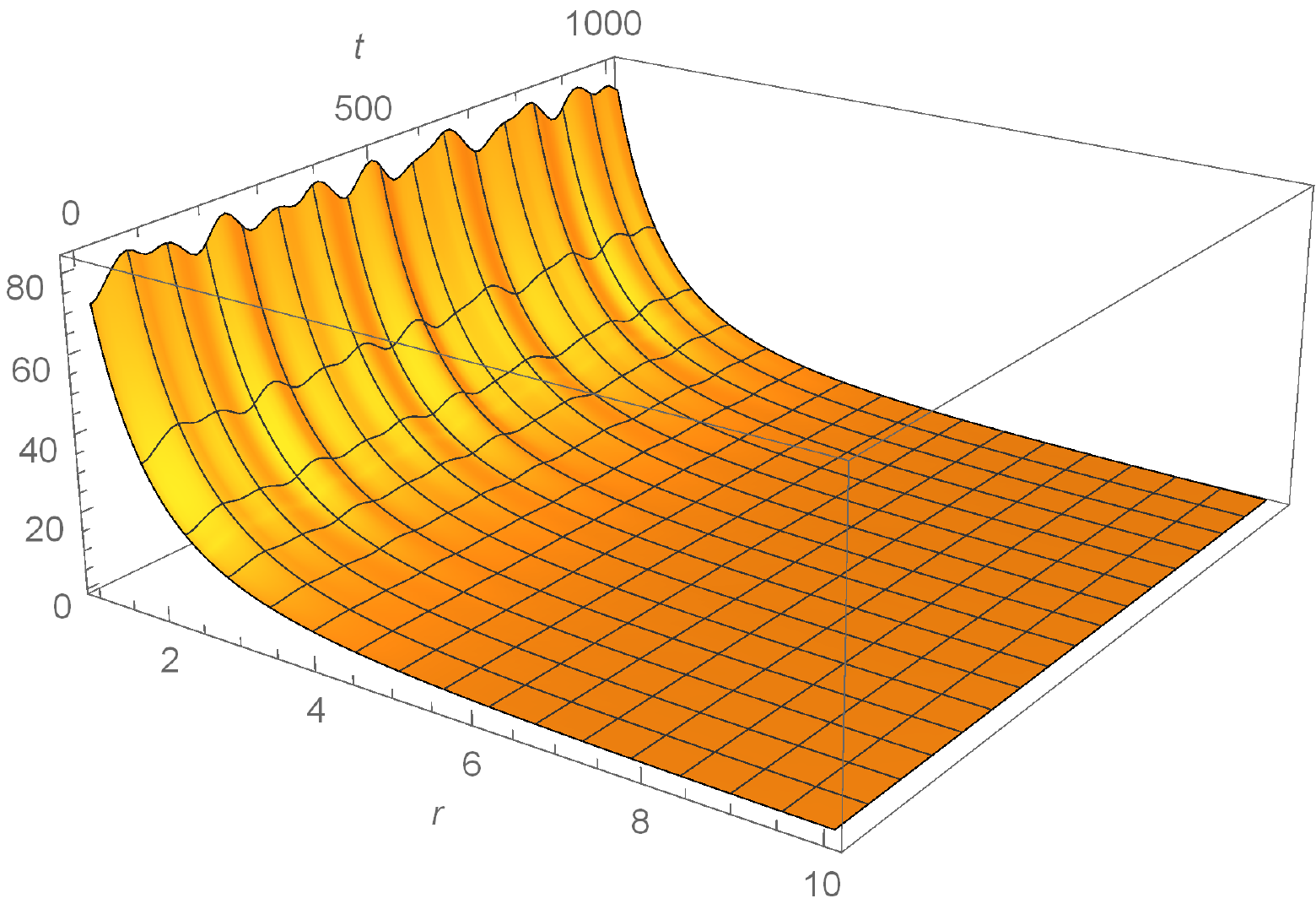}
\includegraphics[width=0.6\columnwidth]{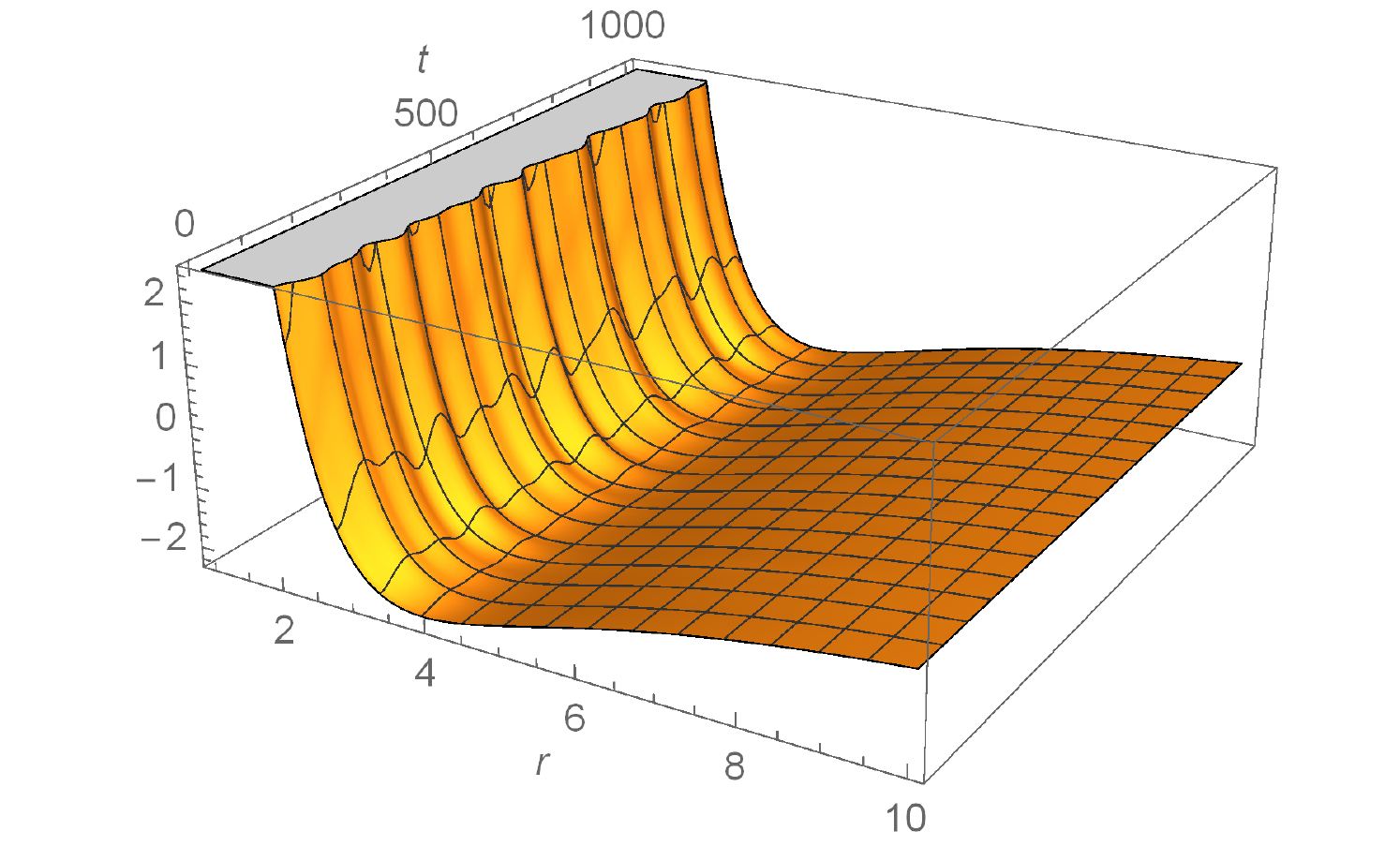}
\caption{Inverse metric components of the first order effective acoustic spacetime :  $g^{tt}_{\text{eff}(1)}(r,t)$ (upper), $g^{tr}_{\text{eff}(1)}(r,t)$ (middle) and $g^{rr}_{\text{eff}(1)}(r,t)$ (lower).}
\end{center}
\end{figure}

In Figure 8, we have plotted an overhead view of $g_{\text{eff}(1)}^{rr}$ and $g^{rr}_{\text{eff}(1)} = 0$ (blue plane) from $r=2.362$ to $r=2.7$ over the entire range of time. The intersection of the two plots traces the horizon radius over time. We see that the horizon fluctuates and grows from the original value $r=2.362$ at $t=0$ to $r=2.468$ at $t=10^3$. At no point after $t=0$ does the horizon ever return to its original value. 
 
\begin{figure}
\begin{center}
\includegraphics[width=1\columnwidth]{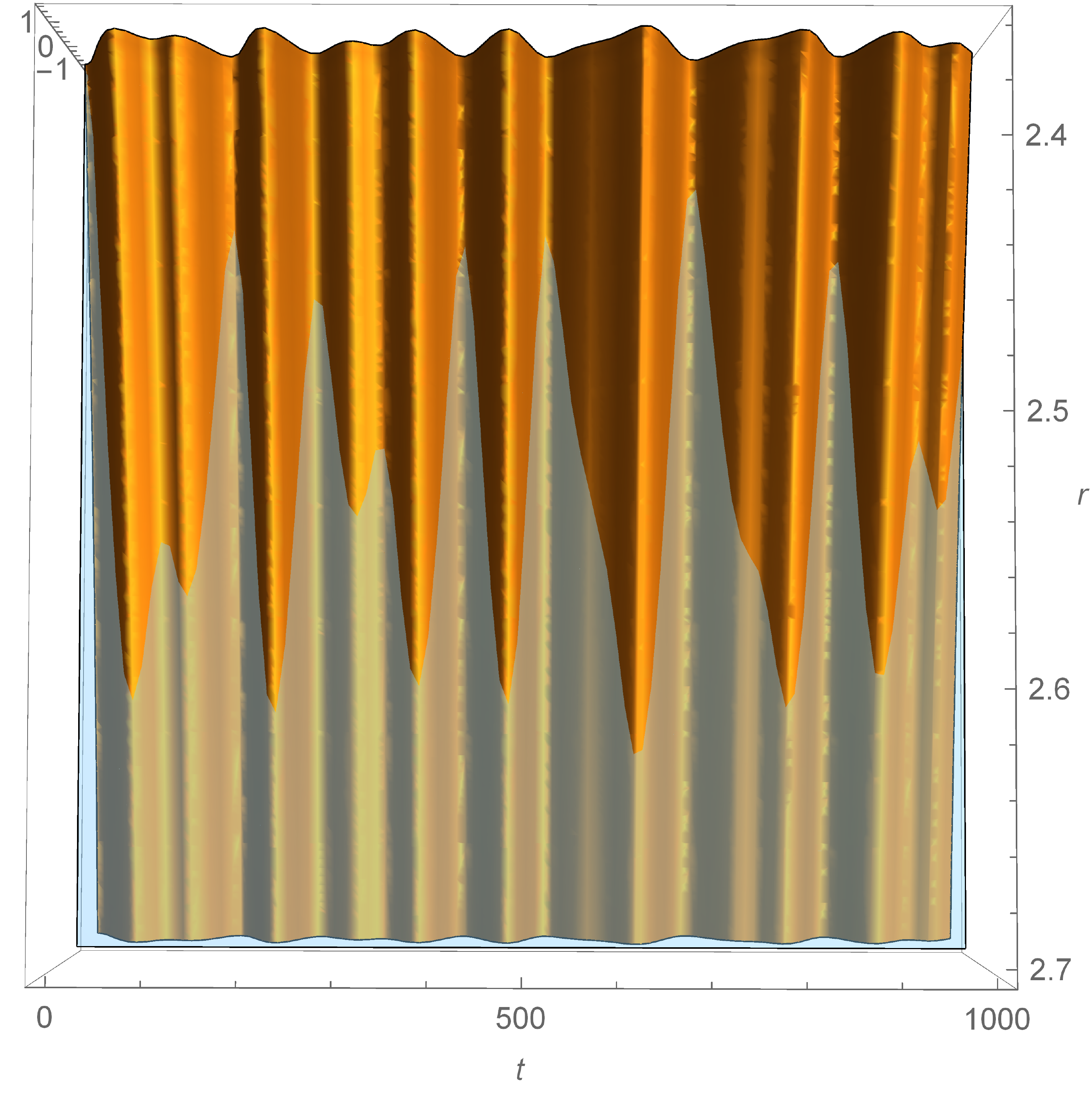}
\caption{Overhead view of $g^{rr}_{\text{eff}(1)}(r,t)$ from $r=2.36$ to $r=2.7$ over all time, from $t=0$ to $t=10^3$. The blue plane represents $g^{rr}_{\text{eff}(1)}(r,t) = 0$ and its intersection with $g^{rr}_{\text{eff}(1)}(r,t)$ traces out the horizon location in time. We note that the horizon fluctuates and grows to a new value of $r=2.468$ at $t= 10^3$.}
\end{center}
\end{figure}

We now proceed to solve the second order perturbation equations for the mass accretion rate and the density given in Eq.~(\ref{r2e}) and Eq.~(\ref{f2e}) respectively. In solving for the mass accretion rate, we provide the initial boundary conditions $f_2(r,0) = f_1(r,0)$ , $f_2(1,t) = f_1(1,t)$  and $f_2(100,t) = f_1(100,t)$, which ensures consistency with the first order perturbation solution and the initial exponentially damped in time behaviour. Likewise for the density perturbation, we require only one condition at initial time $\rho_2(r,0) = \rho_1(r,0)$. The plots of the resulting solution have been provided in Figure 9 and Figure 10. We find that the second order perturbations are more pronounced versions of the first order solutions in the supersonic region. In particular, the density perturbation decays even more at this order as can be seen from comparing Figure 10 with Figure 5. However, there now are less complete cancellations of the constructive and destructive modes of the mass accretion rate fluction $f_2$ in the subsonic region. Hence the corresponding solution for $\rho_2$ also involve certain small corrections that begin to be develop in the subsonic region at this order.  

\begin{figure}
\begin{center}
\includegraphics[width=0.8\columnwidth]{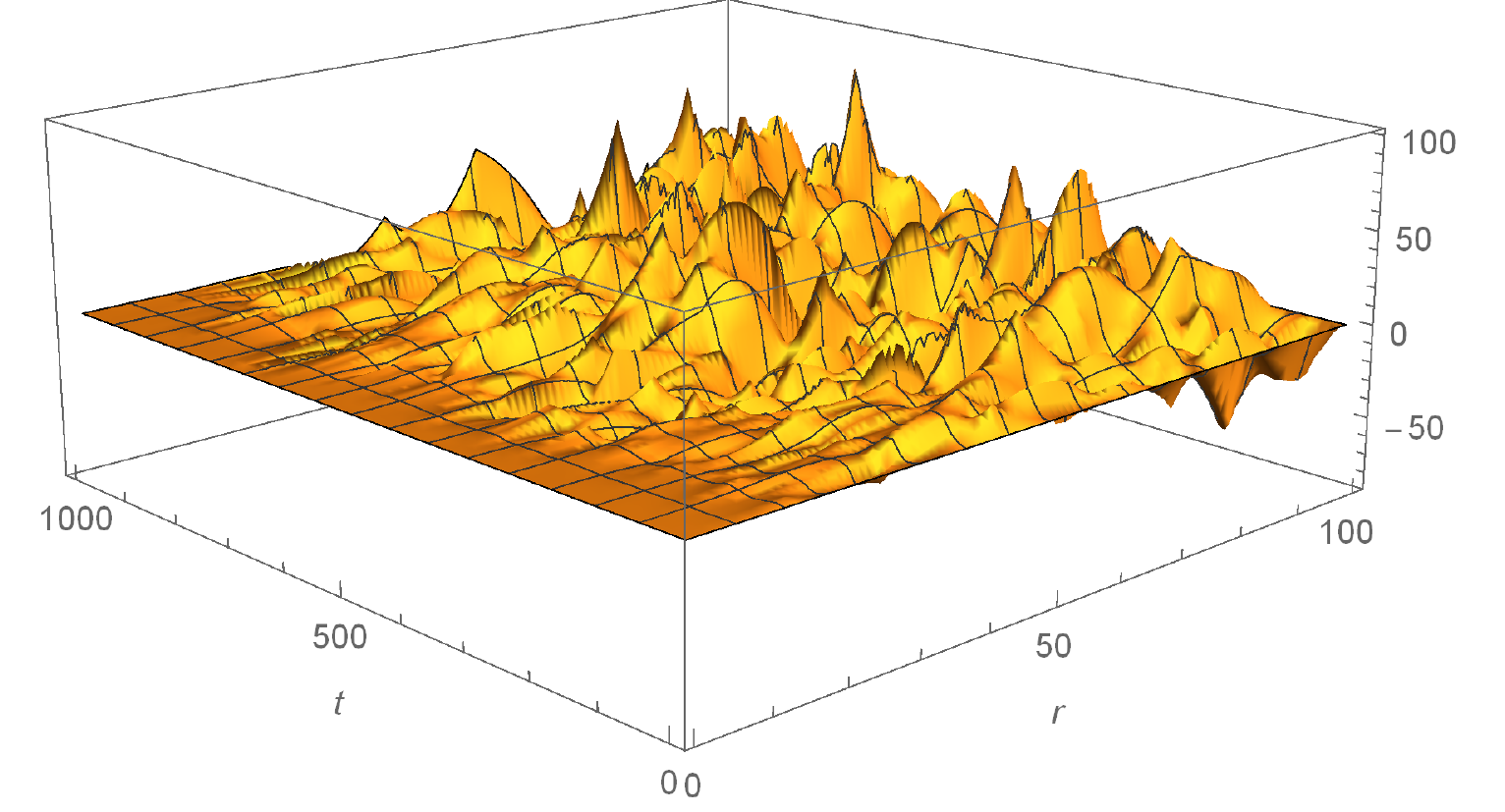}
\caption{Solution of the second order pertubed mass accretion rate $f_2(r,t)$.}
\end{center}
\end{figure}
\begin{figure}
\begin{center}
\includegraphics[width=0.8\columnwidth]{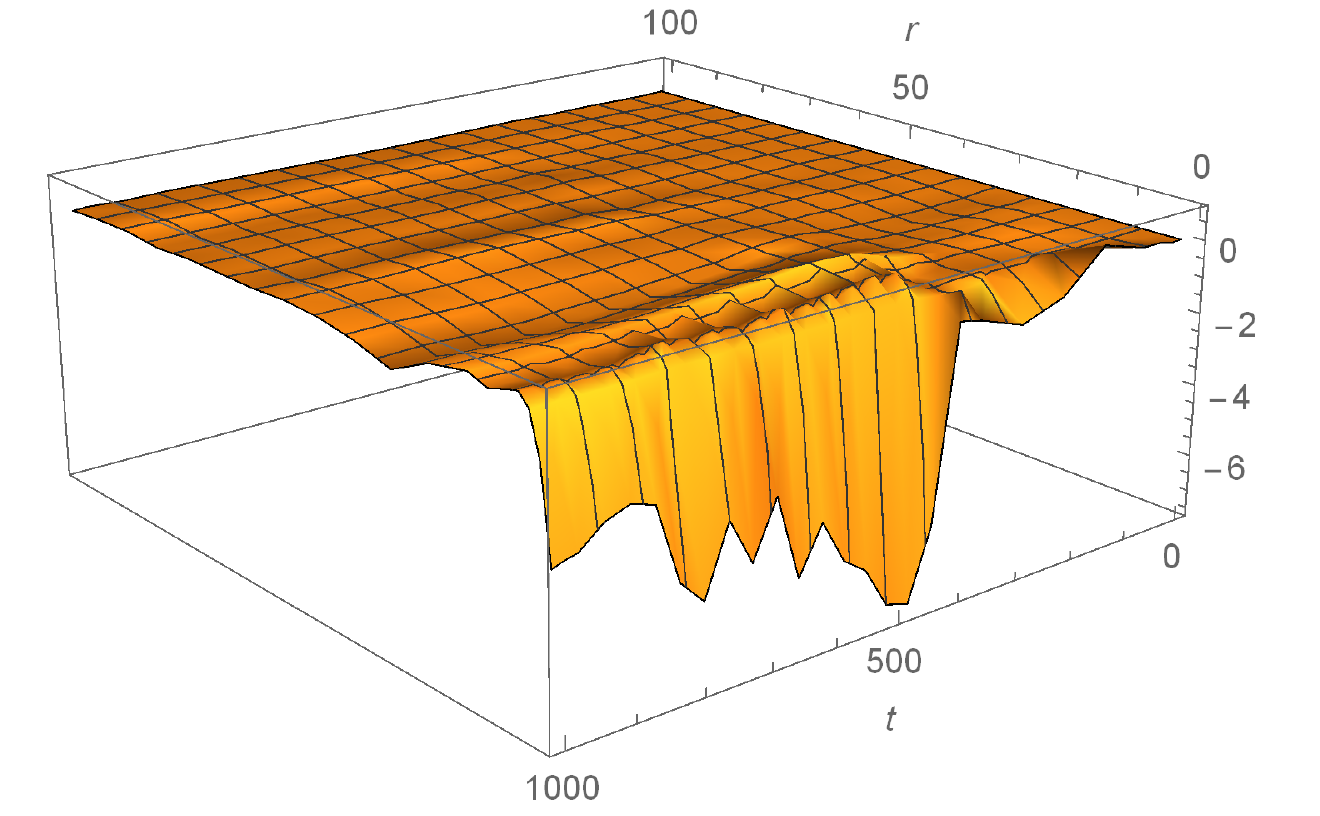}
\caption{Solution of the second order perturbed density $\rho_2(r,t)$. The solution is more negative near the accretor boundary than the corresponding first order solution in Figure 5.}
\end{center}
\end{figure}

With the second order solutions at hand, we can determine the second order effective acoustic metric on which the third order mass accretion rate fluctuation propagates. 
The second order inverse metric components are determined by solving Eq.~(\ref{met2}), which provide the plots in Figure 11. Comparison of these plots with those of the first order solutions in Figure 6 demonstrate that the trends are simply more pronounced at this order, with some small perturbations developing in the subsonic region. The largest magnitude as in the first order perturbation is present in $g^{rr}_{(2)}(r,t)$ at $r=1\,, t=632$. While there is an order of magnitude increase in the maximum value in going from $g^{rr}_{(1)}(r,t)$ to $g^{rr}_{(2)}(r,t)$, the suppression in going from $\epsilon$ to $\epsilon^2$ is more significant. This suppression would remain had we chosen the strength of the perturbation to be closer to $1$ in Eq.~(\ref{npert}).

\begin{figure}
\begin{center}
\includegraphics[width=0.8\columnwidth]{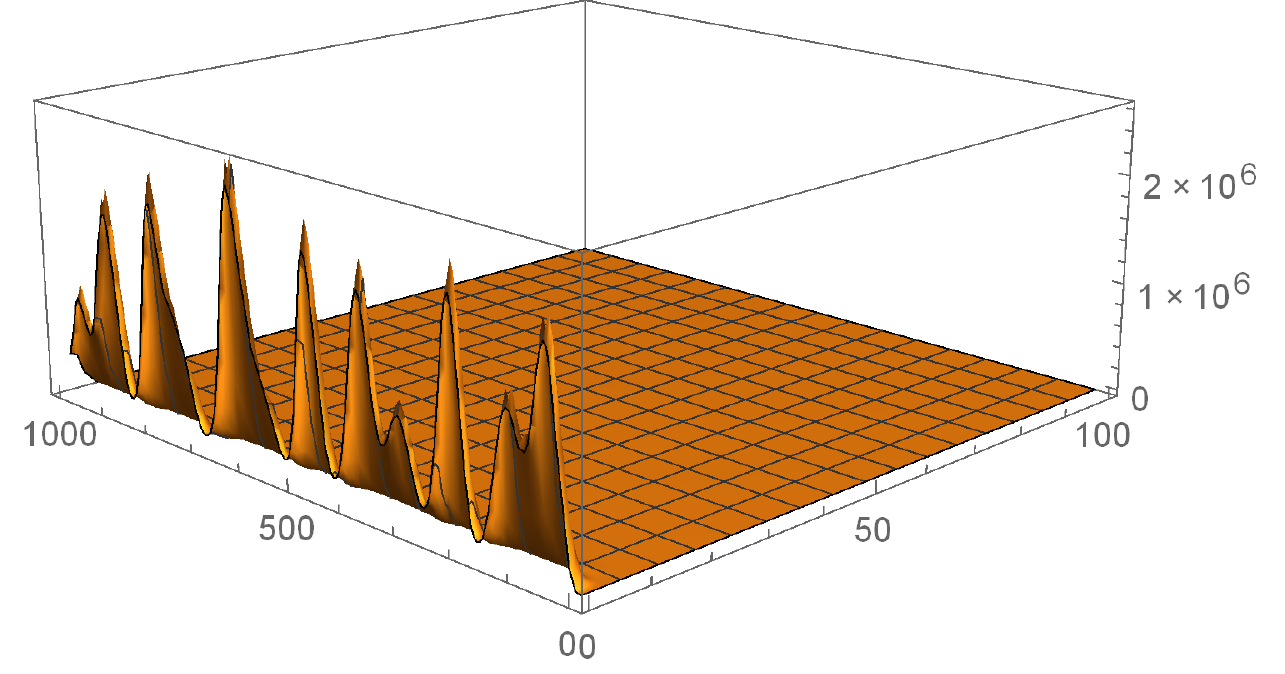}
\includegraphics[width=0.8\columnwidth]{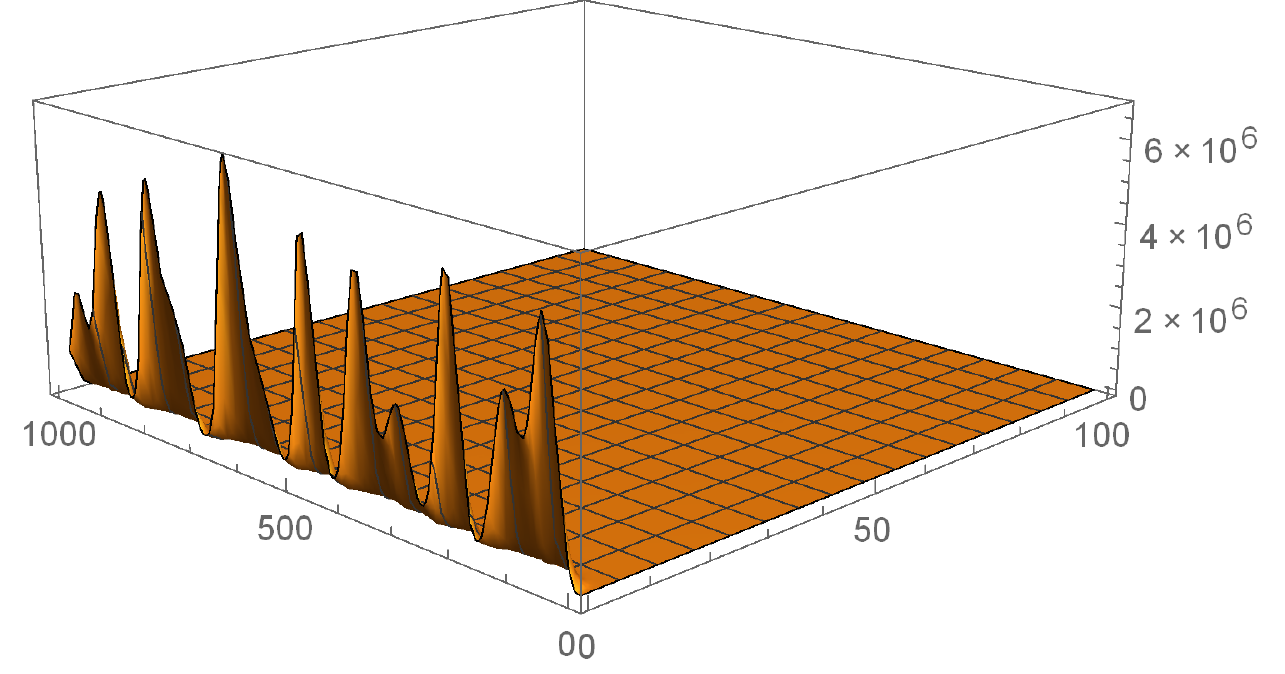}
\includegraphics[width=0.8\columnwidth]{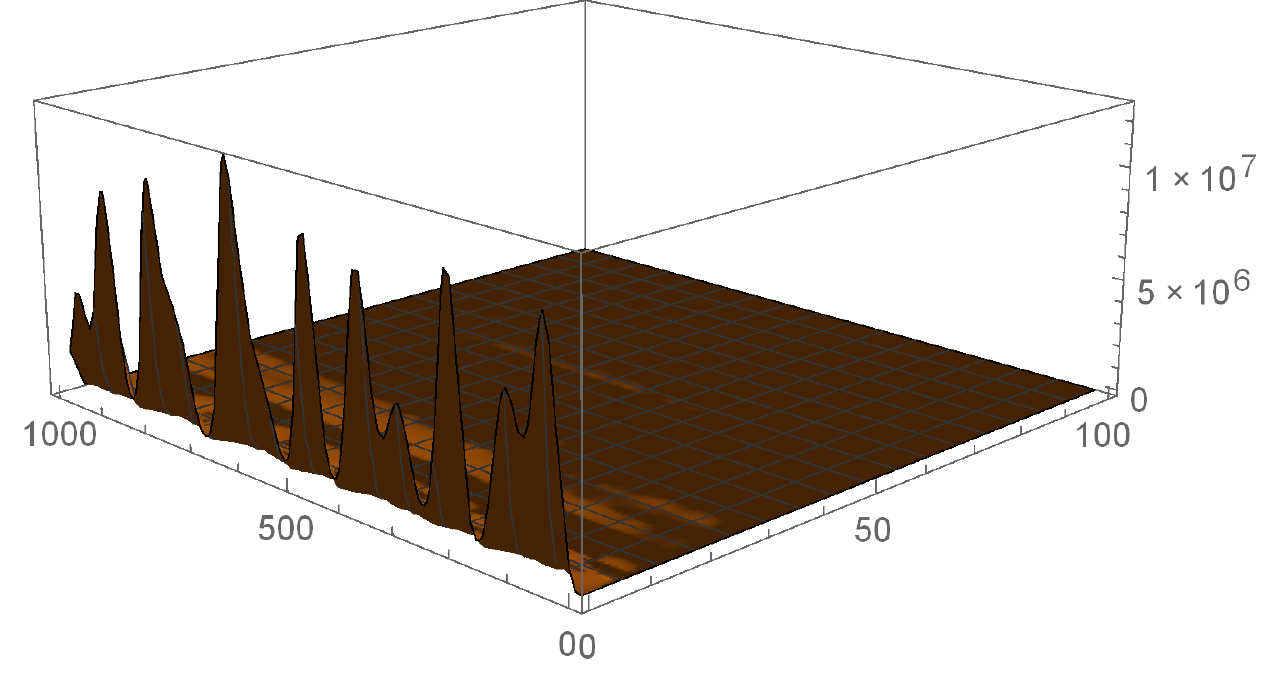}
\caption{Second order perturbed inverse metric components: $g^{tt}_{(2)}(r,t)$ (upper), $g^{tr}_{(2)}(r,t)$ (middle) and $g^{rr}_{(2)}(r,t)$ (lower).}
\end{center}
\end{figure}

We can construct the effective metric up to this order
\begin{equation}
g_{\text{eff}(2)}^{\mu\nu} =  g^{\mu\nu}_{(0)} +  \epsilon g^{\mu\nu}_{(1)} + \epsilon^2 g^{\mu\nu}_{(2)} \,.
\label{nm1.met2}
\end{equation} 

The components of the second order effective acoustic metric are nearly identical to those in Figure 7.  In Figure 12, we have provided the more revealing plot at this order of $g^{rr}_{\text{eff}(2)}(r,t)$ and $g^{rr}_{\text{eff}(2)}(r,t) = 0$ (blue plane) from $r=2.362$ to $r=2.7$ over all time, just as in Figure 8 for $g^{rr}_{\text{eff}(1)}(r,t)$. In comparing with $g^{rr}_{\text{eff}(1)}(r,t)$, we note that the horizon of $g^{rr}_{\text{eff}(2)}(r,t)$ has grown slightly at late times, from $r_H = 2.468$ to $r_H = 2.476$, with no qualitative difference from the first order effective acoustic spacetime result. 

\begin{figure}[H]
\begin{center}
\includegraphics[width=1\columnwidth]{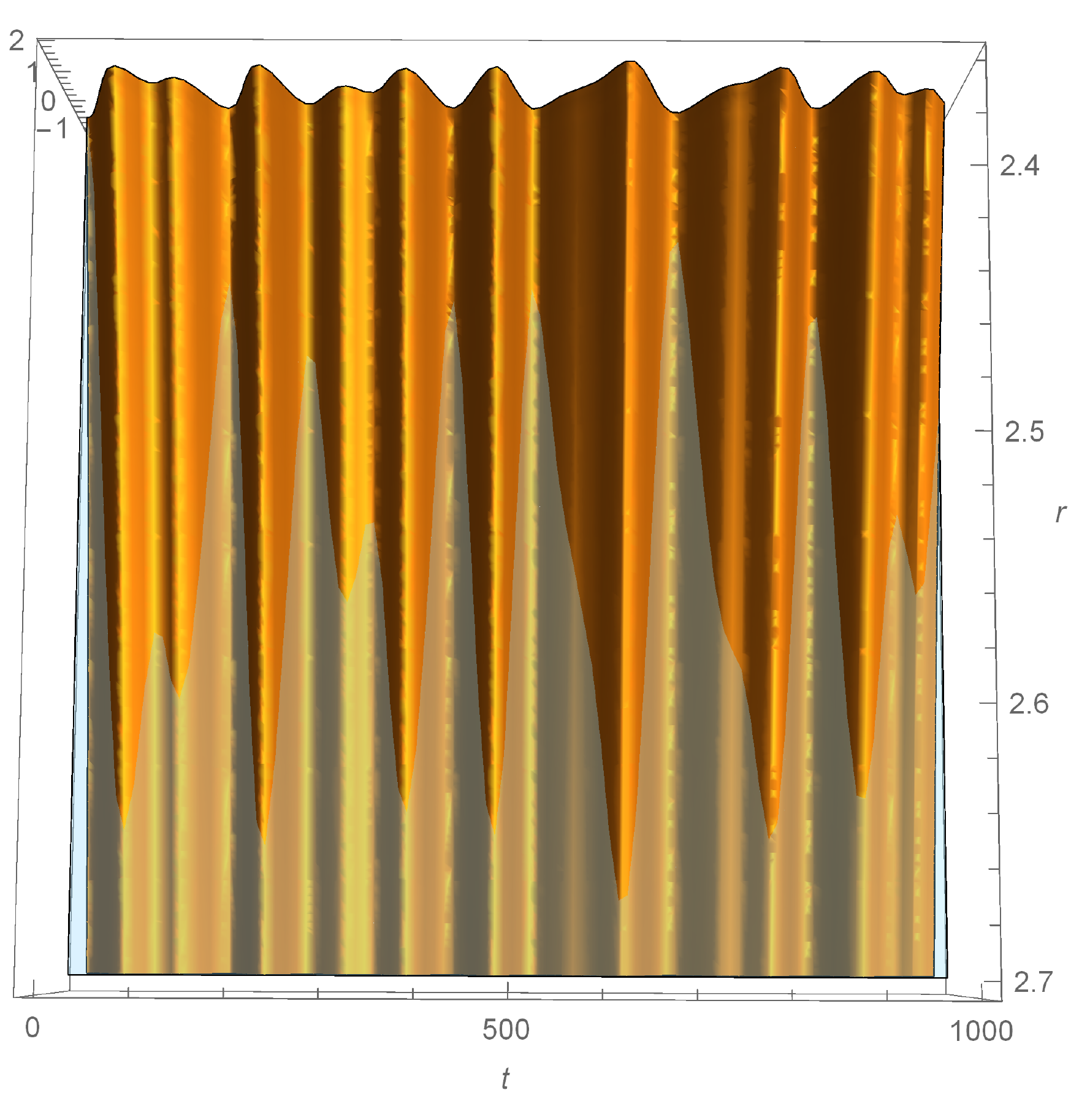}
\caption{$g^{rr}_{\text{eff}(2)}(r,t)$ and $g^{rr}_{\text{eff}(2)} = 0$ from $r=2.362$ to $r=2.7$ over all time. The acoustic horizon has grown slightly relative to $g^{rr}_{\text{eff}(1)}(r,t)$, with $r_H = 2.476$ at $t=10^3$.}
\end{center}
\end{figure}

\subsection{Low frequency perturbations} \label{lfp}
We will now consider low frequency damped perturbations by setting $\omega = \omega_{\text{low}} = 10^{-3}$ for $t \in \{0, 10^3\}$. We solve Eq.~(\ref{wav.1}) with the initial boundary conditions $f_1(r,0) = f_0$ and  $f_1(1,t) = e^{-\omega t} = f_1(100,t)$. The solutions is given in Figure 13. We one again find a solution involving the destructive interference of exponentially growing and decaying waves over space and time in the subsonic region. The only constructive contribution, as in the high frequency case, arises from ingoing waves in the supersonic region. 

The density perturbation $\rho_1$ from solving Eq.~(\ref{r1e}) is plotted in Figure 14. While we expect that the density fluctuation should be significant only in the supersonic region, we find that unlike the corresponding high frequency solution it is positive close to the accretor boundary. This result can be understood from the larger wavelengths of the perturbations in the present case. The perturbations in the low frequency case can neither easily pass through the accretor boundary, nor can they be reflected since their wavelengths are of the order of (or larger than) the accreting fluid. Hence the ingoing perturbations in the supersonic region tend to increase the density close to the accretor leading to the solution as in Figure 14.  

\begin{figure}[H]
\begin{center}
\includegraphics[width=0.8\columnwidth]{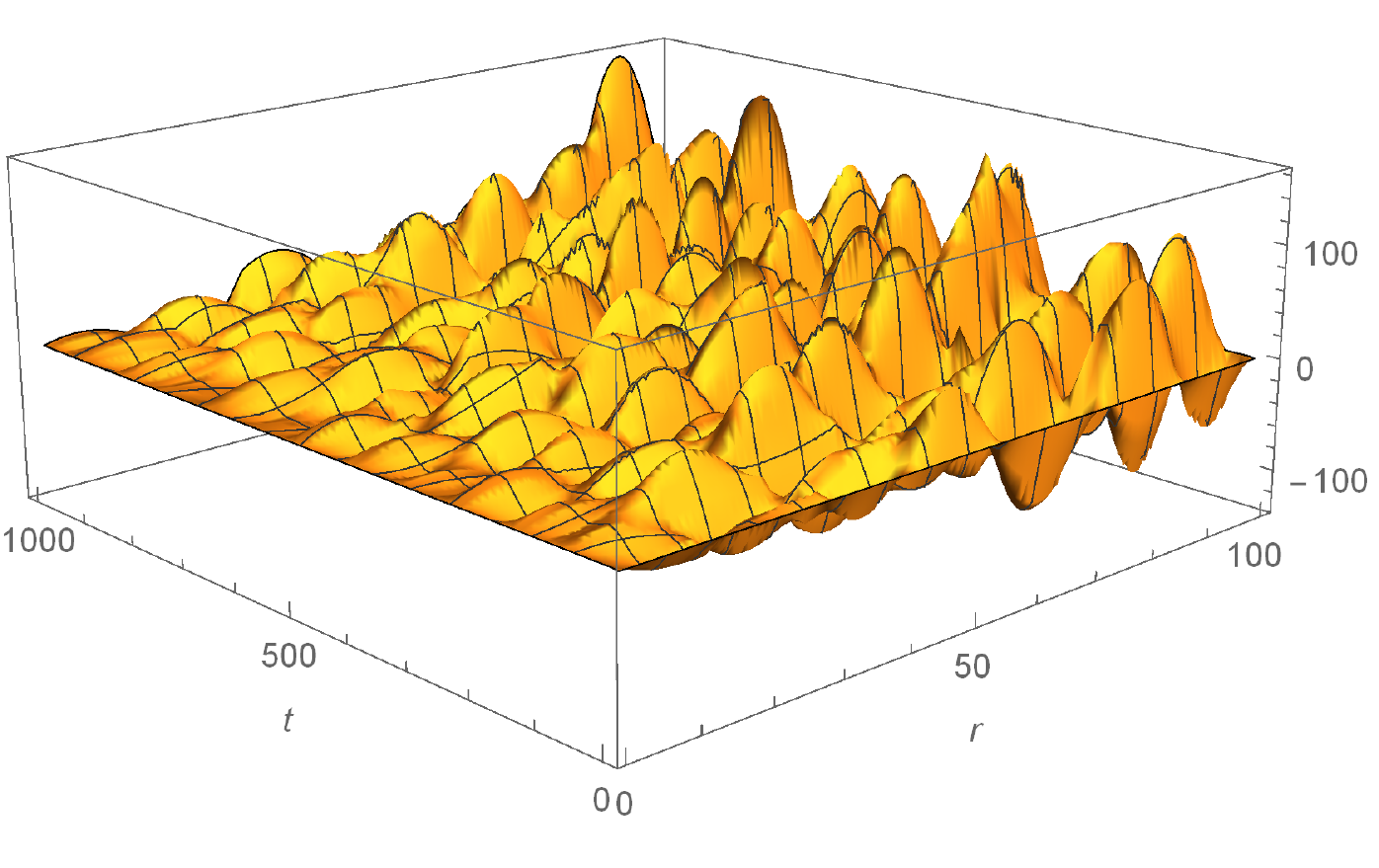}
\caption{Solution of the first order pertubed mass accretion rate $f_1(r,t)$.}
\end{center}
\end{figure}

\begin{figure}
\begin{center}
\includegraphics[width=0.8\columnwidth]{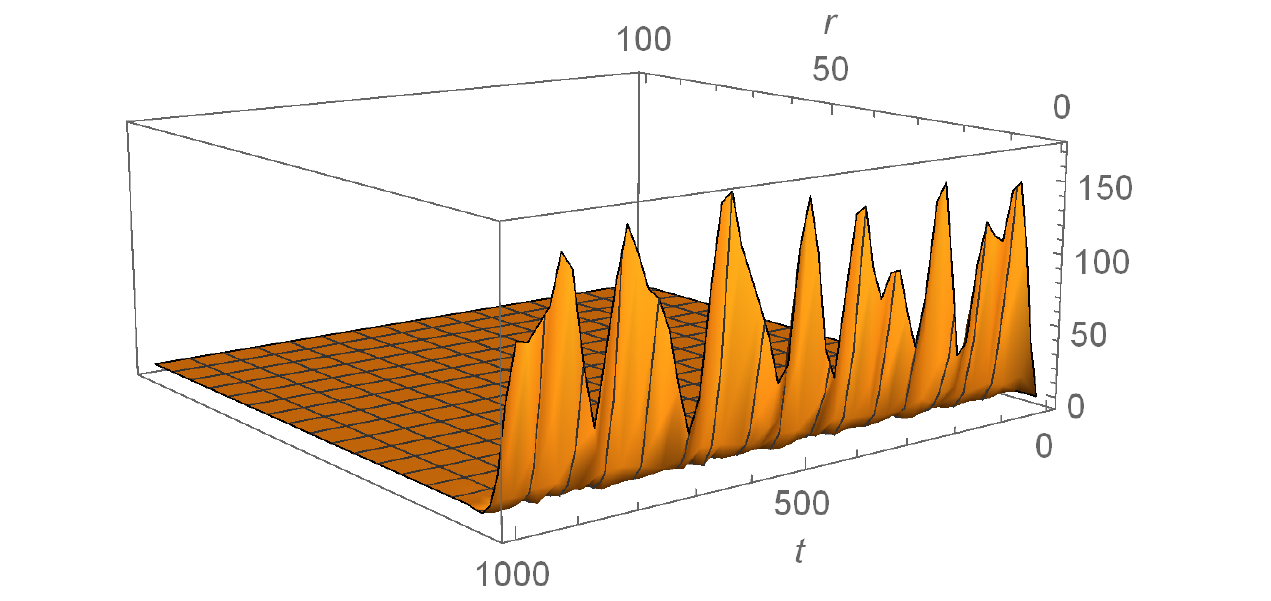}
\caption{Solution of the first order perturbed density $\rho_1(r,t)$. The perturbation is only relevant near the accretor boundary and positive.}
\end{center}
\end{figure}
Since $\rho_1$ is positive near the accretor for low frequency perturbations, this increases the first order corrected sound speed relative to the stationary solution value. It can also be noted from the values of $f_1$ and $\rho_1$ close to the accretor that $\frac{\delta \rho}{\rho_0} > \frac{\delta f}{f_0}$. Hence from the relative fluctuation relation in Eq.~(\ref{varhor}), we expect there to be a receding horizon in this case.

Let us first consider the effect of the first order perturbation solutions for $f_1$ and $\rho_1$ on the inverse metric components in Eq.~(\ref{met1}). The plots of these components are given in Figure 15. As in the high frequency case (plotted in Figure 6), the corrections are only dominant near the accretor boundary. However, unlike the corresponding high frequency solutions, the corrections are negative and are a further indication of the receding acoustic horizon we expect to observe. The maximum absolute value in these plots occurs for $g^{rr}_{(1)}$ at $r=1$ and $t= t_m = 632$. Using Eq.~(\ref{npert}) this sets $\epsilon \sim 6 \times 10^{-6}$ in the low frequency case.

\begin{figure}
\begin{center}
\includegraphics[width=0.8\columnwidth]{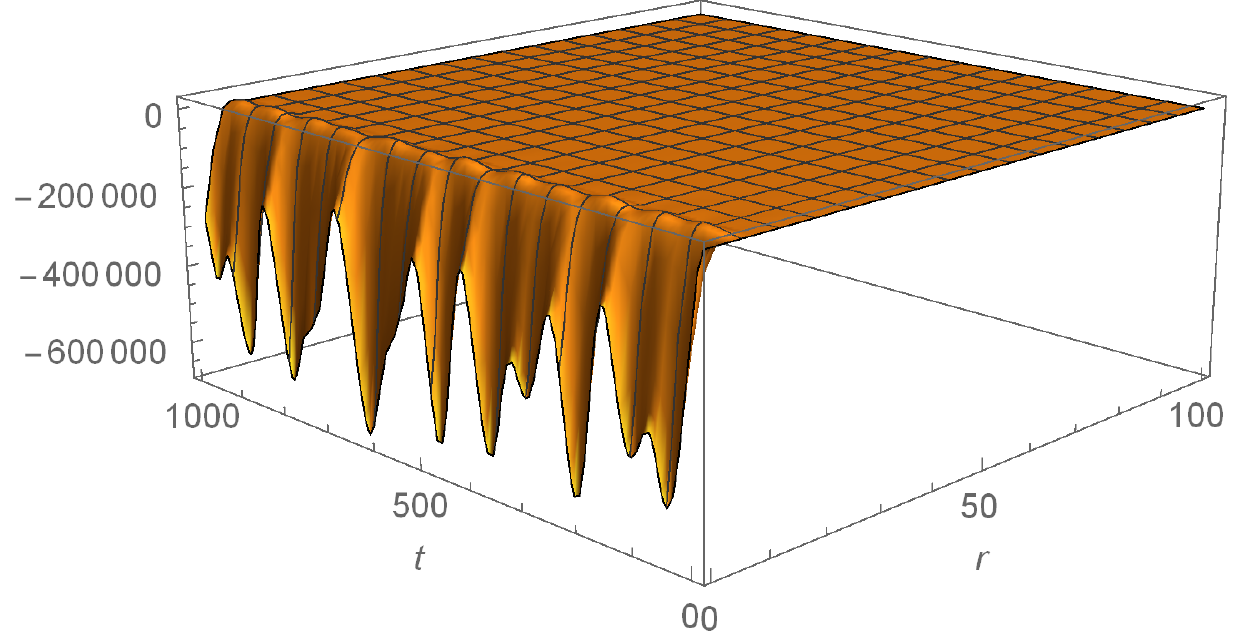}
\includegraphics[width=0.8\columnwidth]{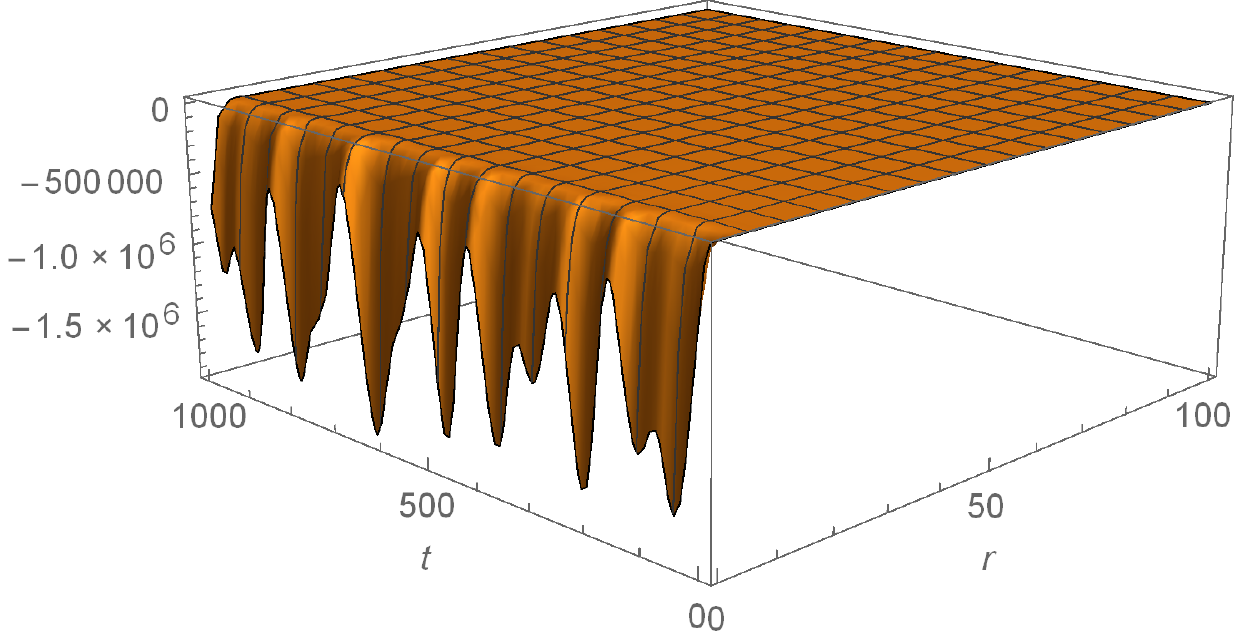}
\includegraphics[width=0.8\columnwidth]{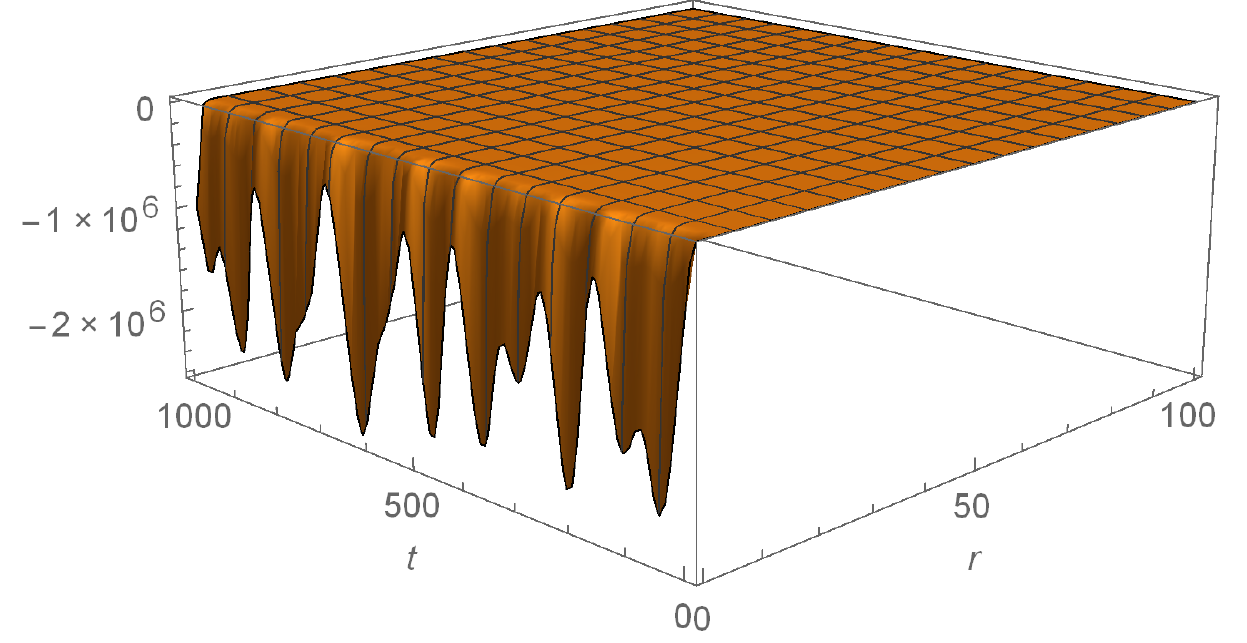}
\caption{First order perturbed inverse metric components: $g^{tt}_{(1)}(r,t)$ (upper), $g^{tr}_{(1)}(r,t)$ (middle) and $g^{rr}_{(1)}(r,t)$ (lower). Unlike the corresponding high frequency perturbation plots in Figure 6, we now find negative corrections near the accretor boundary at $r=1$.}
\end{center}
\end{figure}
We can now use Eq.~(\ref{nm1.met1}) to determine the inverse metric solutions for the first order effective acoustic spacetime. 
As noted in the high frequency analysis, the interesting effects are present in close up plots in the near horizon region. In figure 16, we have plotted $g_{\text{eff}(1)}^{rr}$ and $g_{\text{eff}(1)}^{rr} =0$ (blue plane), now from $r=1.9$ to $r=2.362$ over the entire range of time. The intersection of the two plots traces the horizon radius over time. We see that the horizon fluctuates and shrinks from the original value $r=2.362$ at $t=0$ to $r=2.213$ at $t=10^3$. 

\begin{figure}
\begin{center}
\includegraphics[width=1\columnwidth]{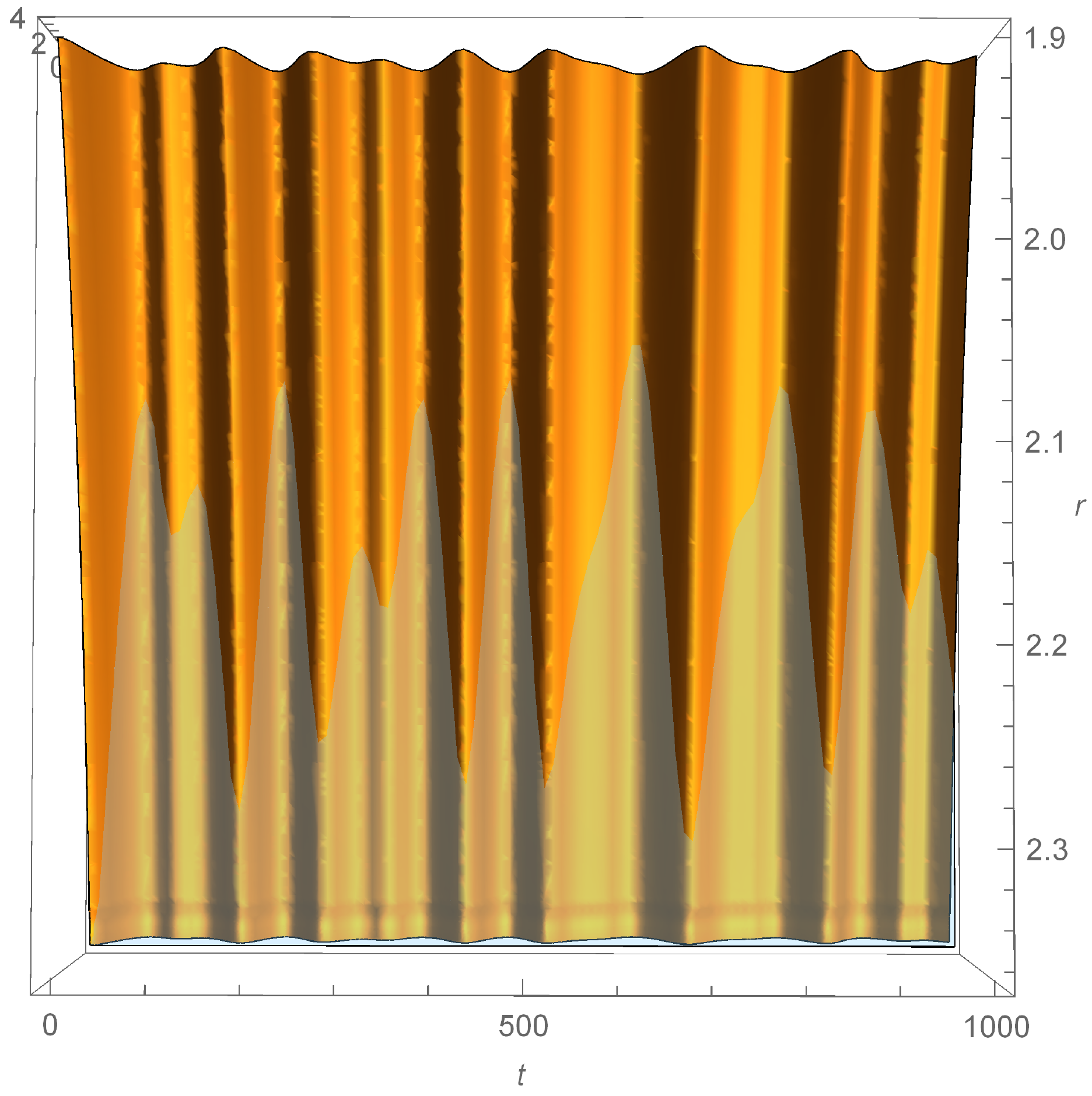}
\caption{$g^{rr}_{\text{eff}(1)}(r,t)$ and $g^{rr}_{\text{eff}(1)} = 0$ from $r=1.9$ to $r=2.3632$ over all time. The horizon fluctuates and shrinks to its new value of $r=2.213$ at $t= 10^3$.}
\end{center}
\end{figure}
We now solve the second order perturbation equations of the mass accretion rate and the density given in Eq.~(\ref{r2e}) and Eq.~(\ref{f2e}) respectively. The initial boundary conditions for the mass accretion rate are chosen to be $f_2(r,0) = f_1(r,0)$ , $f_2(1,t) = f_1(1,t)$  and $f_2(100,t) = f_1(100,t)$, while for the density perturbation we choose $\rho_2(r,0) = \rho_1(r,0)$. The plots of the resulting solutions have been provided in Figure 17 and Figure 18.

In comparing Figure 13 and Figure 17, we find that the mass accretion rate perturbations are approximately the same order of magnitude in going from first order to second order pertubations. The comparison of Figure 14 and Figure 18 on the other hand show that the density perturbation decreases by an order in magnitude going from first order perturbations to those at second order. Hence on the basis of Eq.~(\ref{varhor}), we now expect the acoustic horizon to start growing from this order.

\begin{figure}
\begin{center}
\includegraphics[width=1\columnwidth]{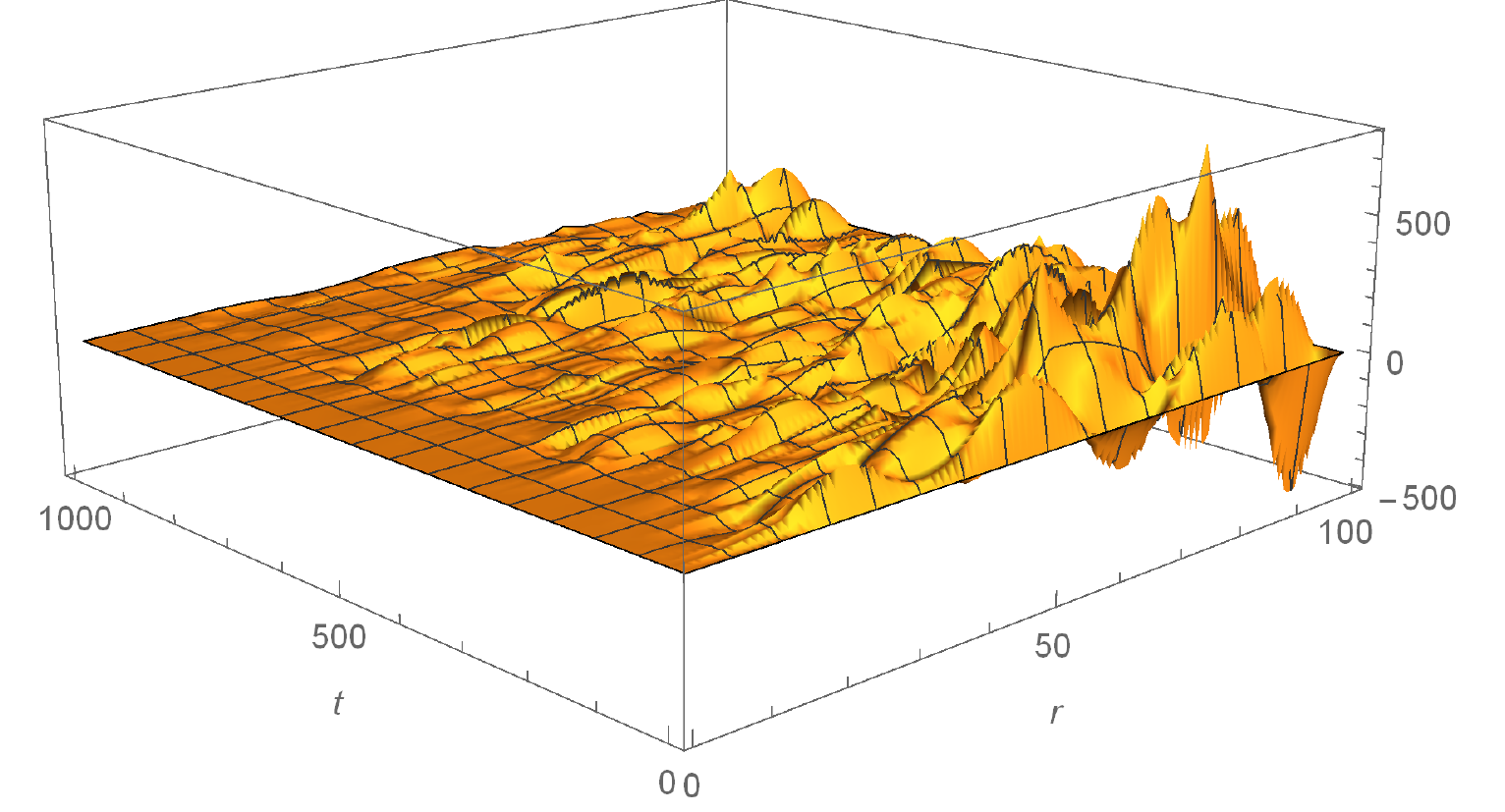}
\caption{Solution of the second order pertubed mass accretion rate $f_2(r,t)$.}
\end{center}
\end{figure}
\begin{figure}
\begin{center}
\includegraphics[width=1\columnwidth]{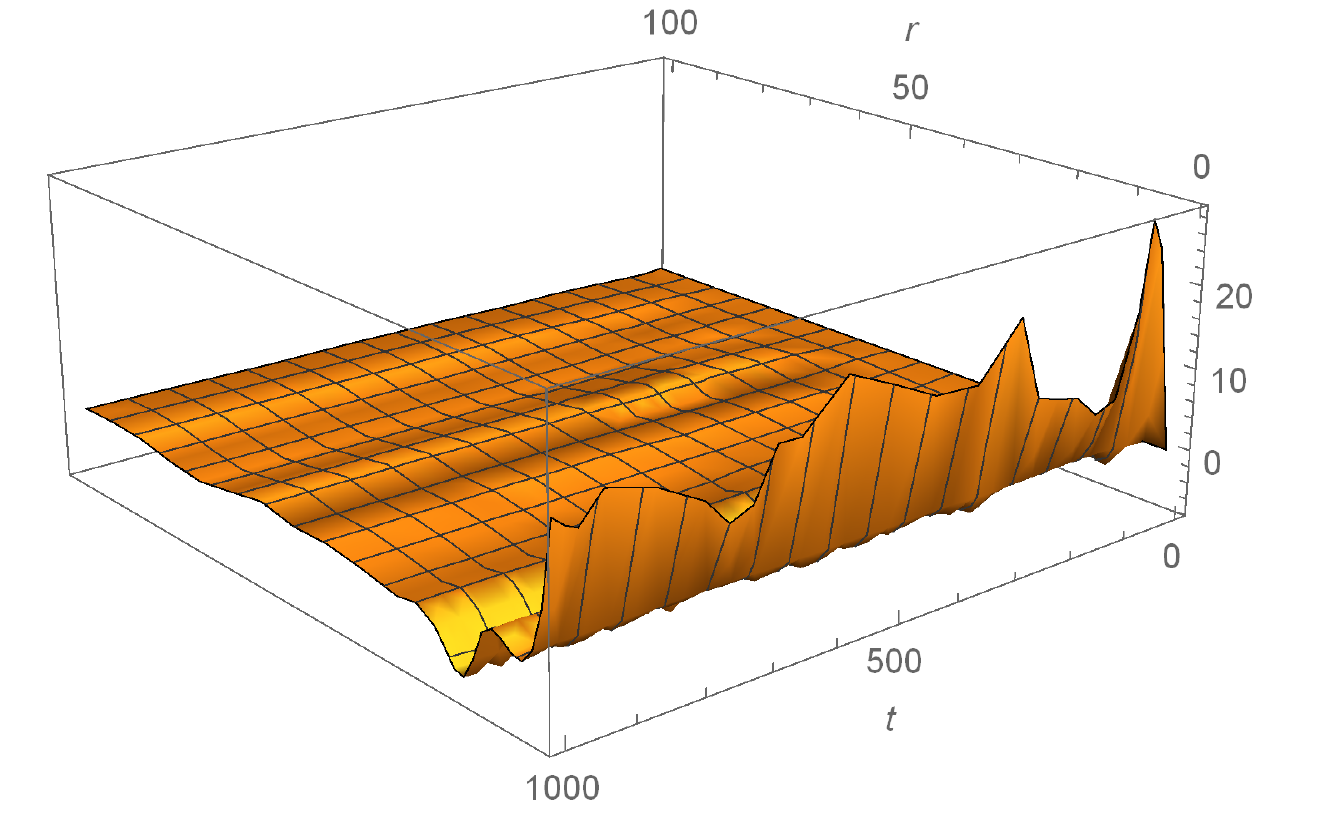}
\caption{The solution of the second order perturbed density $\rho_2(r,t)$. This solution is more negative near the accretor boundary than the first order solution in Figure 15.}
\end{center}
\end{figure}

The second order inverse metric components of Eq.~(\ref{met2}) are plotted in Figure 19. At second order, these components are qualitatively the same as the high frequency solutions in Figure 6 and Figure 11. This further supports the expected relative increase in the horizon size in going to the second order effective acoustic spacetime. However just as in the high frequency case, we expect this increase to be marginal since the suppresion from higher powers of $\epsilon$ dominate over the growth of the perturbed inverse metric components, regardless of the strength of the perturbation. 
In Figure 20, we have plotted $g_{\text{eff}(2)}^{rr}$ and $g_{\text{eff}(2)}^{rr} =0$ (blue plane) from $r=1.9$ to $r=2.362$ over the entire time range. The horizon radius in going from first to second order effective acoustic spacetime increases from $r_H=2.213$ to $r_H = 2.228$ at $t=10^3$. Thus the result of a receding horizon persists at second order. 

\begin{figure}
\begin{center}
\includegraphics[width=0.8\columnwidth]{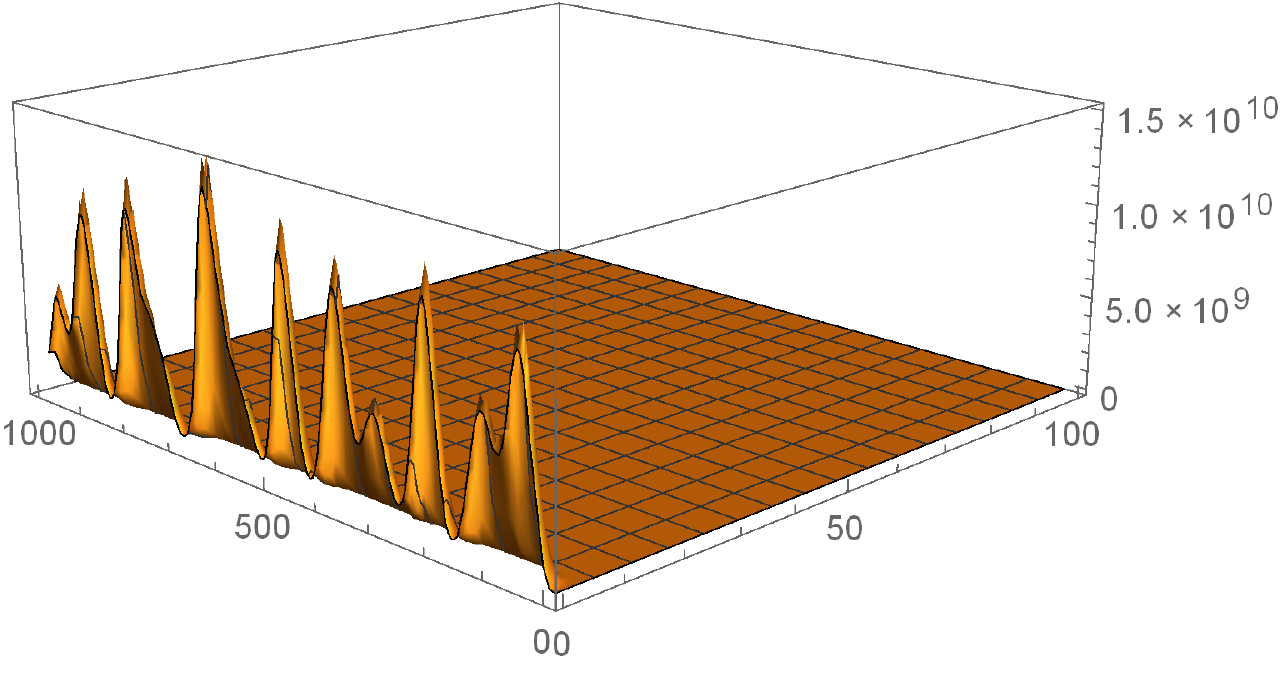}
\includegraphics[width=0.8\columnwidth]{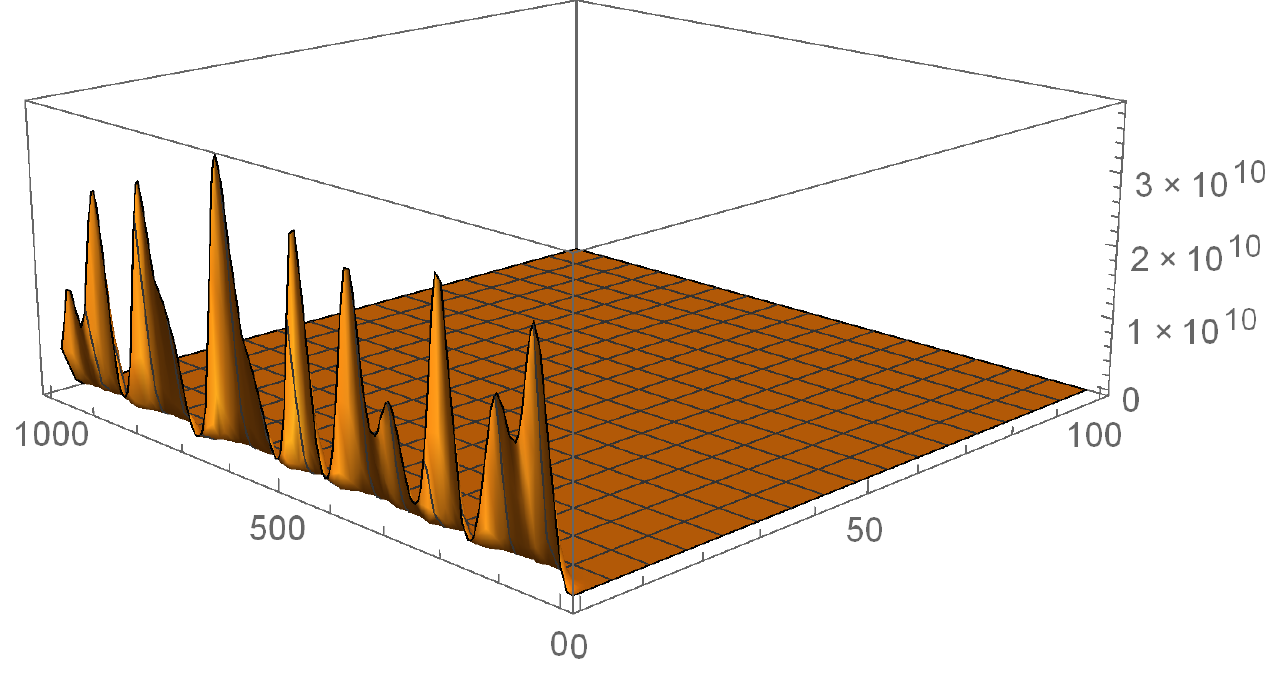}
\includegraphics[width=0.8\columnwidth]{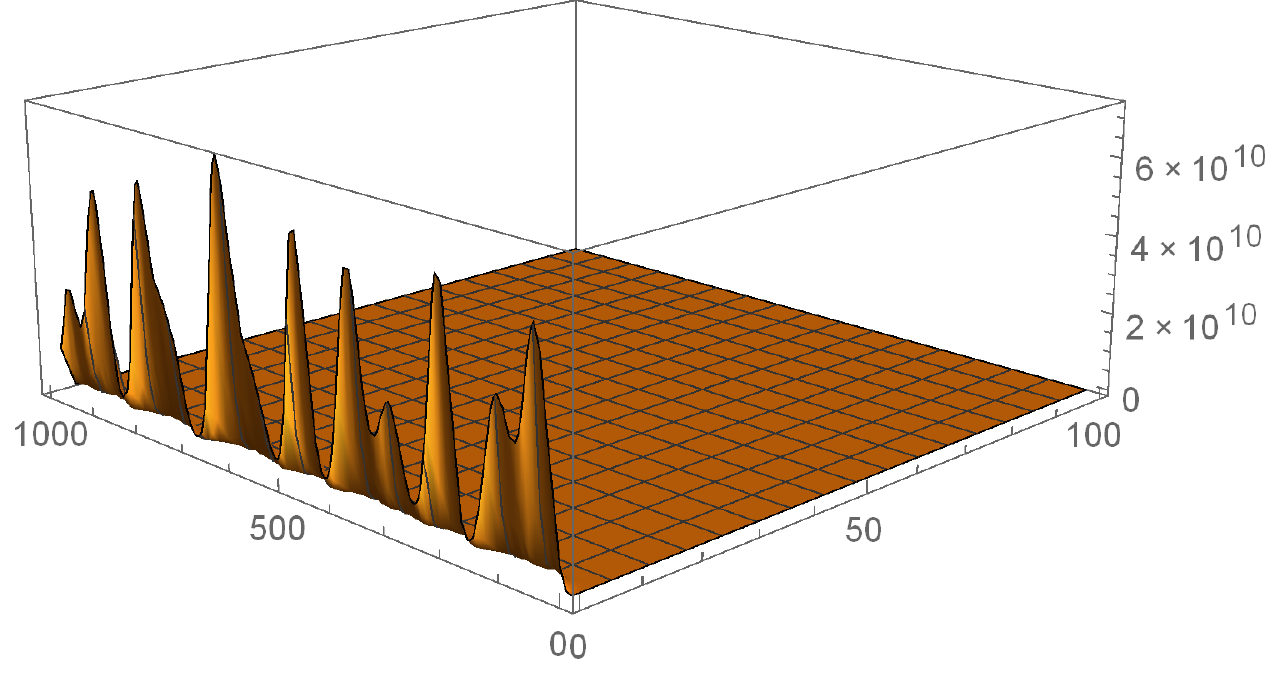}
\caption{Second order perturbed inverse metric components: $g^{tt}_{(2)}(r,t)$ (upper), $g^{tr}_{(2)}(r,t)$ (middle) and $g^{rr}_{(2)}(r,t)$ (lower). Unlike the first order solutions in Figure 16, these solutions take on the qualitative form of high frequency solutions in Figure 6 and Figure 11.}
\end{center}
\end{figure}

\begin{figure}
\begin{center}
\includegraphics[width=0.8\columnwidth]{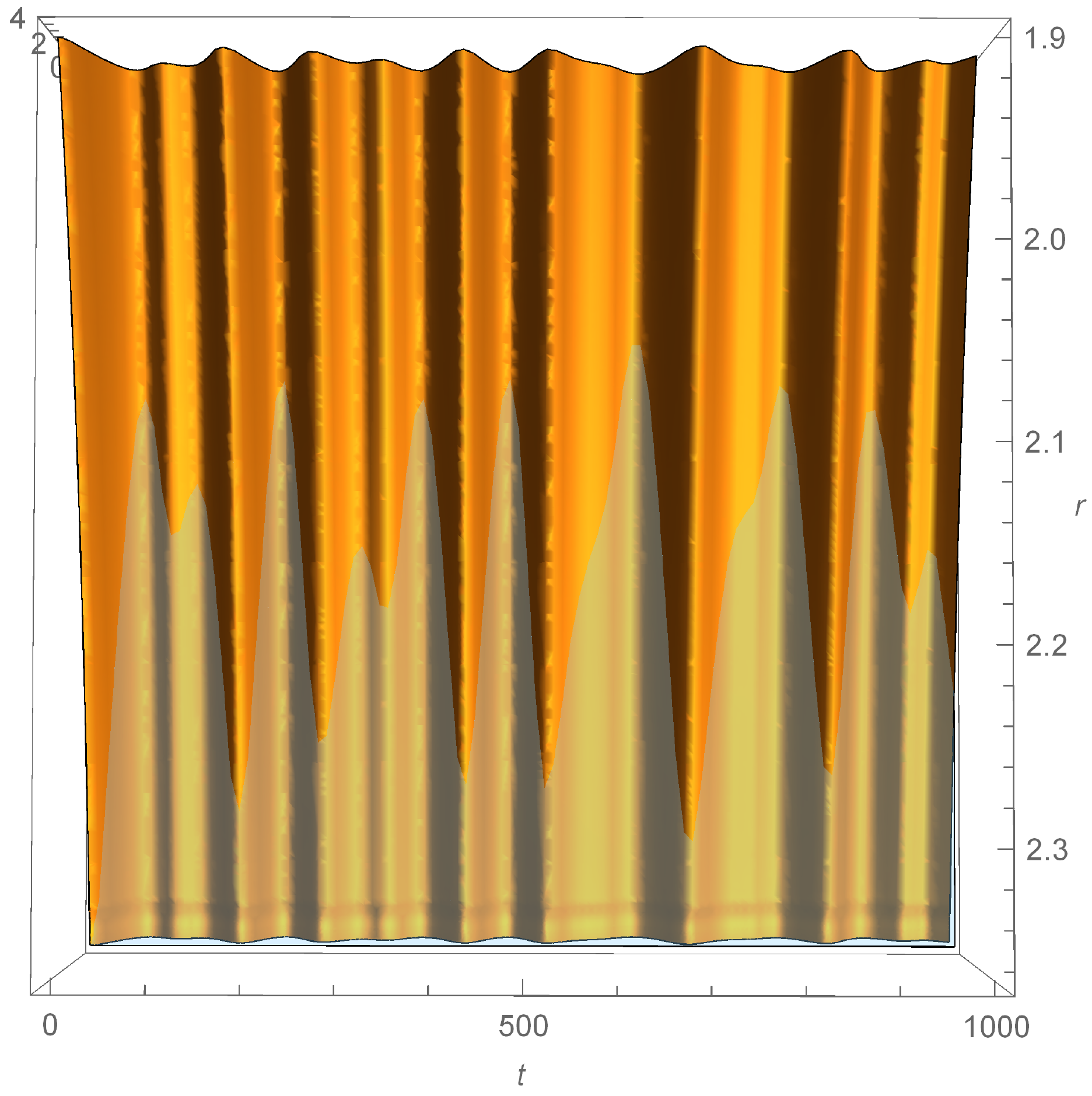}
\caption{$g^{rr}_{\text{eff}(2)}(r,t)$ and $g^{rr}_{\text{eff}(2)} = 0$ from $r=1.9$ to $r=2.362$ over all time. The second order effective acoustic horizon has grown slightly more than than in $g^{rr}_{\text{eff}(1)}(r,t)$ at $t=10^3$ given in Figure 17, from $r=2.213$ to $r=2.228$.}
\end{center}
\end{figure}
In conclusion, we see that in the case of very low frequency damped perturbations, there is the interesting effect of a fluctuating acoustic horizon that recedes in size. This effect is well approximated to higher orders in perturbation, owing to the suppression from higher powers of the parameter $\epsilon$ regardless of the perturbation strength.

\section{Discussion} \label{disc}

In this paper, we introduced a formalism to investigate dynamical analogue spacetimes from nonlinear perturbations of transonic fluids. In Sec.~\ref{nrf}, we considered a class of spherically symmetric non-relativistic flows described in terms of the mass accretion rate and the density of the fluid. This was shown to provide a continuity equation relating the mass accretion rate and density, and a wave equation for the mass accretion rate. In particular, the wave equation is a direct consequence of considering the mass accretion rate as an independent variable and does not generically result from other variable choices. We then demonstrated that the perturbative expansion of the independent variables leads to a wave equation for the fluctuating mass accretion rate at a given order, propagating on an effective dynamical analogue spacetime constructed from lower order solutions. The continuity equation allows for the fluctuation in density to be determined entirely from the fluctuation in the mass accretion rate to all orders in perturbation. Hence, the perturbative formalism provides an iterative scheme to derive nonlinear perturbations of transonic fluids and thereby dynamical analogue spacetimes to all orders..  

In Sec.~\ref{curve} we investigated properties of these dynamical analogue spacetimes by interpreting the `background' in Eq.~(\ref{wav}) as its closed form expression from all perturbation orders. We demonstrated that given finite density and mass accretion rate values throughout the flow, there exists a well defined casual structure and curvature to all orders in perturbation. We also determined an expression for the relative fluctuation of the effective acoustic horizon in terms of relative fluctuations of the mass accretion rate and density. Using this relation, we identified that the acoustic horizon can recede under classical perturbations, unlike black holes, in scenarios where the density fluctuation in the supersonic region is positive and grows more than the mass accretion rate fluctuation. 

In Sec.~\ref{num}, we numerically investigated the effective acoustic spacetime to second order in perturbations of the Bondi accreting flow solution. This analysis was performed on time scales comparable to the size of the accretion flow and in two distinct regimes of `High frequency' and `Low frequency' perturbations. Our analysis involved the choice of an exponentially damped in time perturbation throughout the fluid, with appropriately chosen boundary conditions. The case of high frequency perturbations involve wavelengths smaller than the size of the accretor. The perturbations in this case were shown to have the effect of increasing the mass accretion rate and decreasing the density of the fluid across the acoustic horizon, to all orders in perturbation. This leads to an acoustic horizon that fluctuates and grows to a larger radius at late times, analogous to the ringdown ot black holes under perturbations.

Conversely, low frequency perturbations involved wavelengths larger than the size of the fluid. This frequency regime could be defined due to the presence of a maximum length scale associated with the system, namely the scale of the fluid itself. Such perturbations are also of potential importance in astrophysical accretion flows and could provide a novel probe in experimental tests of analogue spacetimes. To linear order, we noted that low frequency perturbations have the property of being trapped in the supersonic region close to the accretor boundary. In the example of Bondi accretion, the increase in the density fluctuation exceeded the increase in the mass accretion rate, leading to a manifestation of the receding horizon scenario.
The first order effective acoustic spacetime constructed from these solutions is time dependent with an acoustic horizon that fluctuates and shrinks to a new size at late times. At higher orders, fluctuations of the mass accretion rate and density are qualitatively similar to the high frequency case, thereby promoting the growth of the acoustic horizon. However, as shown through both high and low frequency cases, the strength of the perturbation does not significantly modify the near horizon results of the first order effective acoustic spacetime. In particular, the low frequency observation of a receding horizon is not an artefact of perturbation strength and is generically realized. 

Our results can be directly applied to other experimentally accessible transonic flows, such as those realized in de Laval nozzles and shallow water systems.  By considering the mass accretion rate, our approach could also be applied to transonic flows that violate the irrotational condition since there exists no criteria on the velocity. Thus for cases where the fluid velocity involve more than one non-vanishing component, the procedure can in principle be carried out following the definition of mass accretion rates with respect to the velocity components. In addition, from linear perturbations of relativistic accretion flows it is known the mass accretion rate can be taken as the perturbed quantity~\cite{Ananda:2014gga, Ananda:2014qba} and hence the formalism should be readily applicable to relativistic accretion flows. We claim that as the mass accretion rate accounts for the compressibility and can be defined for all fluids, it is likely to prove useful in investigating dynamical aspects of analogue spacetimes from nonlinear perturbations. 

As noted in the introduction, an important application of our formalism will be in the domain of astrophysical accretion flows. Realistic accretion flows are typically axisymmetric and can involve additional effects, in particular the presence of magnetic fields and shocks. In the case of accreting axisymmetric flows including shocks, linear perturbations are known to propagate on analogue spacetimes that feature analogue black holes and white holes~~\cite{Das:2003py,Abraham:2005ah,Das:2006an}. This analogue spacetime structure is a result of multi-transonic flows that are characteristic of axisymmetric accretion flows. The analogue spacetime in such cases will be relevant in describing the terminal behaviour of fluid variables close to the event horizon of the accreting black hole. A fully dynamical description of these analogue spacetimes could also involve interesting effects resulting from the merger of analogue black holes and white holes. Such processes in the analogue spacetime might in particular be associated with phase transitions in accreting flows. We look forward to generalizing our formalism to a broader class of transonic fluids and exploring these applications to accretion flows in the future.

\end{document}